\documentclass[aps,prd,11pt,superscriptaddress,amssymb,amsmath,nofootinbib]{revtex4-1}

\usepackage{hhline}
\usepackage{color}
\usepackage{epsfig}
\usepackage{graphicx}
\usepackage{float}
\usepackage{epstopdf}
\usepackage{hyperref}
\usepackage[usenames,dvipsnames]{xcolor}
\usepackage{multirow}
\usepackage[normalem]{ulem}
\usepackage[d]{esvect}
\usepackage[misc,geometry]{ifsym}

\begin{document}

 \title{Connect the Lorentz Violation  to the Glashow Resonance Event}
 \author{Ding-Hui Xu}\email{3036602895@qq.com}
 \author{Shu-Jun Rong\Letter}\email{rongshj@glut.edu.cn}
 \affiliation{College of Science, Guilin University of Technology, Guilin, Guangxi 541004, China}

 \begin{abstract}
The recent reported Glashow resonance (GR) event shows a promising prospect in the test of the Lorentz violation (LV) by the high-energy astrophysical neutrinos (HANs)
around the resonant energy. However, since the production source and the energy spectra of HANs are uncertain at present,
moderate LV effects may be concealed at the TeV energy-scale. In this paper, we propose the LV Hamiltonian of a special texture which can lead to the decoupling of
$\nu_{\mu}$ ($\bar{\nu}_{\mu}$). On the base of the decoupling, a noticeable damping of the GR event rate is shown for the HANs from the source dominated by $\bar{\nu}_{\mu}$,
irrespective of the energy spectra of the HANs at Earth. Accordingly, the observation of GR events may bring stringent constraints on the LV and the production mechanism of HANs.

\end{abstract}

 \maketitle

 \section{Introduction}

The Glashow resonance (GR)\cite{1} at $E_{\nu}\simeq6.3$ PeV in the rest frame of electrons is a  probe of high-energy astrophysical antineutrinos $\bar{\nu}_{e}$\cite{2,3}.
The observation of  GR events can provide useful information on the energy spectra, production source, and flavor transition of high-energy astrophysical neutrinos (HANs).
In recent years, several PeV events of HANs have been reported by the IceCube neutrino observatory\cite{4}. In particular, the first GR event was detected at the $2.3\sigma$ level\cite{5}.
Although there is just one event, the source where no antineutrinos are produced may be excluded as the sole origin of HANs.
Furthermore, the detection of the GR event can also set constraints on new physics effects which become significant at the PeV energy-scale.
Many novel effects in the field of HANs, such as  those from pseudo-Dirac neutrinos\cite{6,7,8,9}, neutrino-decay\cite{10,11,12,13,14,15}, neutrino secret interactions\cite{16,17,18},
etc\cite{19,20,21,22}, have been proposed. In this paper, we take into account the impacts of the Lorentz violation (LV)\cite{23,24,25,26,27,28,29,30,31,32} on
the flux of $\bar{\nu}_{e}$ around the resonance energy and show an interesting observation that the GR events may be damped by the LV.

As is known, the $\bar{\nu}_{e}$ flux controlling the GR event rate is affected by neutrino oscillations over cosmological distances.
The effects of the LV on neutrino oscillations are determined by the Hamiltonian $H=H_{0}+H_{LV}$, where the first and the second term are expressed respectively as following\cite{33}:
\begin{equation}
\label{eq:1}
H_{0}=
U\left(
    \begin{array}{ccc}
      0 & 0 & 0 \\
      0 & \frac{\Delta m^{2}_{21}}{2E_{\nu}} & 0 \\
      0 & 0 & \frac{\Delta m^{2}_{31}}{2E_{\nu}} \\
    \end{array}
  \right)U^{+},
\end{equation}
\begin{equation}
\label{eq:2}
H_{LV}=\pm\left(
                   \begin{array}{ccc}
                     0 & a^{T}_{e\mu} & a^{T}_{e\tau} \\
                     (a^{T}_{e\mu})^{*} & 0 & a^{T}_{\mu\tau} \\
                      (a^{T}_{e\tau})^{*} &  (a^{T}_{\mu\tau})^{*} & 0 \\
                   \end{array}
                 \right)-\frac{4E_{\nu}}{3}\left(
                   \begin{array}{ccc}
                     0 & c^{TT}_{e\mu} & c^{TT}_{e\tau} \\
                     (c^{TT}_{e\mu})^{*} & 0 & c^{TT}_{\mu\tau} \\
                      (c^{TT}_{e\tau})^{*} &  (c^{TT}_{\mu\tau})^{*} & 0 \\
                   \end{array}
                 \right).
\end{equation}
Here $U$ is the leptonic mixing matrix in vacuum \cite{34} and matter effects are not considered. For antineutrinos, all terms should be replaced by their complex conjugates and the signal of the $a$ matrix is minus. According to the
recent constraints on the LV parameters from neutrino oscillation experiments\cite{33,35}, the energy-scale of the $a$ matrix can take $10^{-24}$GeV at the $95\%$ confidence limit.
Hence, employing the global fit data on the squared mass differences of neutrinos \cite{36}, we can see that $H_{LV}$  becomes dominant in $H$ when the neutrino energy is of order TeV.
Moreover, even if the $a$ ($c$) parameters are as tiny as $10^{-28}$ GeV ($10^{-31}$), $H_{LV}$ is still significant when $E_{\nu}$ reaches PeV.
Accordingly, the $\bar{\nu}_{e}$ flux leading to the GR events is sensitive to the LV effect which cannot be detected by the neutrino oscillation experiments at low energies.

In the light of the realistic situation of neutrino astronomy, to obtain information on LV from the GR events, two troublesome issues should be addressed.
First, although the first promising point-source of HANs, namely the blazar TXS0506+056, was identified at the $3\sigma$ level\cite{37}, the genuine discovery of the origin of HANs is still in lack at present epoch.
Thus, the original flavor composition of the HANs at the source is totally unknown. For the sake of undermining this problem, two typical sources, namely the pion ($\pi^{\pm}$) decay source, the muon ($\mu^{\pm}$) damped
source, are considered  in this paper. Second, the fitted energy spectra of HANs by now have large uncertainties.
Several models, such as the single power-law, the double power-law,  the power-law with cut-off, etc, can coordinate the observed events of HANs\cite{4,38}.
Furthermore, the constraints on the parameters of the spectra are loose. Thus, moderate LV effects on GR events may be concealed by the uncertainties of the energy spectra.
To overcome this issue, we propose the matrix $H_{LV}$ of a special texture which can  bring the decoupling of $\nu_{\mu}$ ($\bar{\nu}_{\mu}$).
The GR events from the typical sources are damped by the decoupling effect.
Consequently, stringent constraints on the LV Hamiltonian and the production source of HANs may be obtained.

In the following section, the specific impacts of the LV on the flavor transition of HANs, the flux of $\bar{\nu}_{e}$, and the strength of the GR event rate are shown. Finally, conclusions are given.

\section{Glashow resonance impacted by Lorentz violation}

\subsection{The flavor transition of HANs and the texture of  $H_{LV}$}
The coherence of neutrino mass eigenstates  has disappeared after HANs travelled a cosmological distance. Hence, the averaged neutrino oscillation probability is written as
 \begin{equation}
\label{eq:3}
\overline{P}_{\alpha\beta}^{s}=\sum_{i}|U_{\alpha i}|^{2}|U_{\beta i}|^{2},
\end{equation}
where $\alpha, \beta = e, \mu, \tau$, $i = 1, 2, 3$.
This is the so-called  standard flavor transition probability.
When the LV is taken into account, $\overline{P}_{\alpha\beta}^{s}$ is changed into $\overline{P}_{\alpha\beta}$ which reads
\begin{equation}
\label{eq:4}
\overline{P}_{\alpha\beta}=\sum_{i}|U^{N}_{\alpha i}|^{2}|U^{N}_{\beta i}|^{2}.
\end{equation}
The new mixing matrix $U^{N}$ is obtained by the diagonalization of the total Hamiltonian $H=H_{0}+H_{LV}$.
The eigenvalues of $H$ are written as\cite{33}
\begin{equation}
\label{eq:5}
E_{i}=-2\sqrt{Q}\cos(\frac{\theta_{i}}{3})-\frac{a}{3},~~i=1,~2,~3.
\end{equation}
The components are expressed as follows\cite{33}:
\begin{equation}
\label{eq:6}
Q=\frac{a^{2}-3b}{9},
\end{equation}
\begin{equation}
\label{eq:7}
\theta_{1}=\arccos(RQ^{-\frac{3}{2}}),~~\theta_{2}=\theta_{1}+2\pi,~~\theta_{3}=\theta_{1}-2\pi,
\end{equation}
\begin{equation}
\label{eq:8}
a=-Tr(H),~~b=\frac{Tr(H)^{2}-Tr(H^{2})}{2},~~ c=-det(H),
\end{equation}
\begin{equation}
\label{eq:9}
R=\frac{2a^{3}-9ab+27c}{54},
\end{equation}
where the notations Tr and det denote the trace and determinant respectively.
Accordingly, the elements of $U^{N}$ are written as \cite{33}
\begin{equation}
\label{eq:10}
U^{N}_{ei}=\frac{B_{i}^{*}C_{i}}{N_{i}},~~U^{N}_{\mu i}=\frac{A_{i}C_{i}}{N_{i}},~~U^{N}_{\tau i}=\frac{A_{i}B_{i}}{N_{i}},
\end{equation}
in which
\begin{equation}
\label{eq:11}
A_{i}=H_{\mu\tau}(H_{ee}-E_{i})-H_{\mu e}H_{e\tau},
\end{equation}
\begin{equation}
\label{eq:12}
B_{i}=H_{\tau e}(H_{\mu\mu}-E_{i})-H_{\tau\mu }H_{\mu e},
\end{equation}
\begin{equation}
\label{eq:13}
C_{i}=H_{\mu e}(H_{\tau\tau}-E_{i})-H_{\mu\tau}H_{\tau e},
\end{equation}
\begin{equation}
\label{eq:14}
N_{i}^{2}=|A_{i}B_{i}|^{2}+|A_{i}C_{i}|^{2}+|B_{i}C_{i}|^{2}.
\end{equation}

Given the analytical expression of $U^{N}$ and the specific LV parameters, we can predict how the flavor ratio at the source is changed into that at Earth.
However, the issue of the flavor transition including the LV effects is more subtle. On the one hand, since the probability $\overline{P}_{\alpha\beta}$ is dependent on both the magnitude
and the texture of the matrix $H_{LV}$, too many patterns of $\overline{P}$ can be obtained from  the LV parameters in the allowed ranges shown in Tab.\ref{tab:1}.
On the other hand, the experimental constraints on the flavor ratio of HANs at the PeV energy-scale are loose at present. In order to give robust observations on the impacts of the LV on the GR events,
we propose a special pattern with the decoupling of $\nu_{\mu}$ ($\bar{\nu}_{\mu}$), i.e.,
\begin{equation}
\label{eq:15}
\overline{P}\sim\left(
                  \begin{array}{ccc}
                    \frac{1}{2} & 0 & \frac{1}{2} \\
                    0 & 1 & 0 \\
                     \frac{1}{2}  & 0 &  \frac{1}{2}  \\
                  \end{array}
                \right).
\end{equation}

\begin{figure}
  \centering
  \includegraphics[width=.49\textwidth]{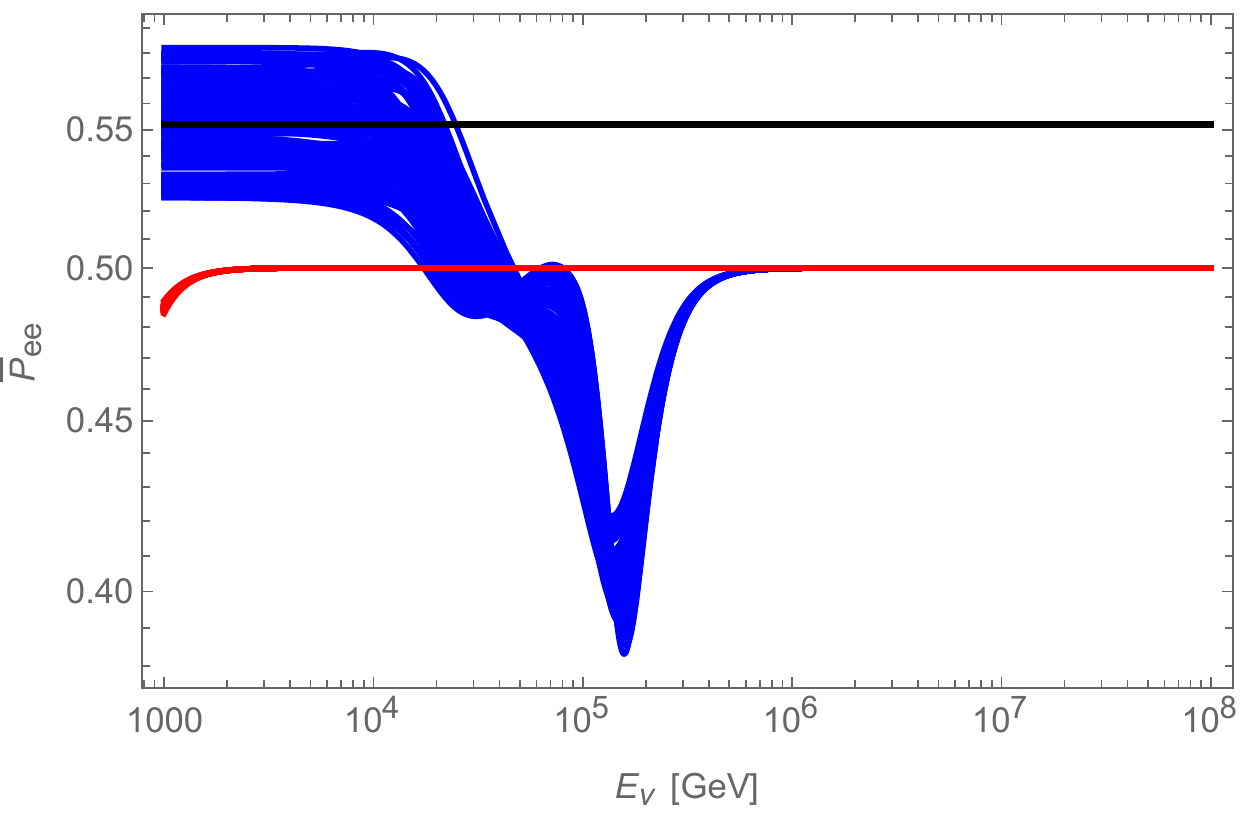}
  \hfill
  \includegraphics[width=.49\textwidth]{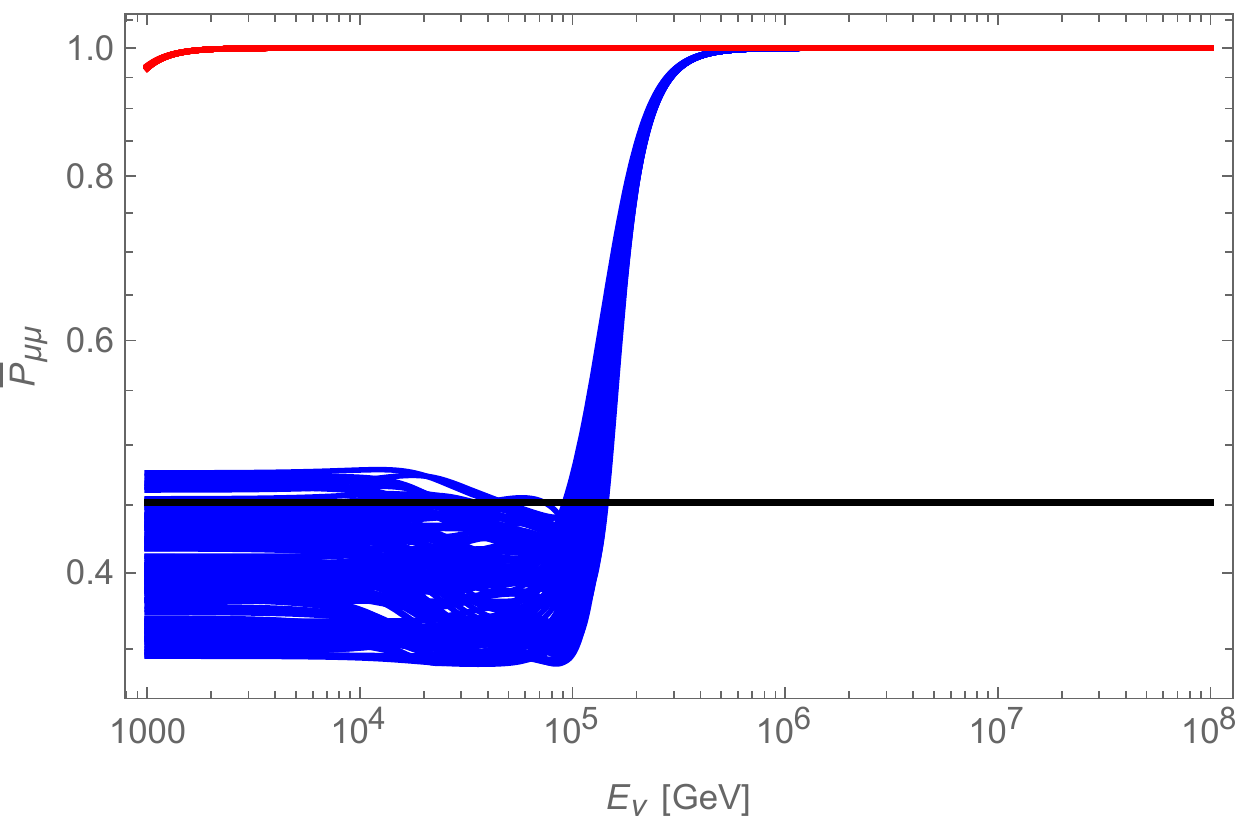}
  \caption{\label{fig:1} The flavor transition probability of antineutrinos.
$\overline{P}_{\alpha\beta}$ is obtained from the leptonic mixing parameters at the $3\sigma$ level of the global fit data with normal
  mass ordering(NO)\cite{36} and the LV parameters.
 For blue lines, $a^{T}_{\alpha\beta}=10^{-7}(a^{T}_{\alpha\beta})_{bf}$, $c^{TT}_{\alpha\beta}=10^{-7}(c^{TT}_{\alpha\beta})_{bf}$. For red lines, $a^{T}_{\alpha\beta}=10^{-3}(a^{T}_{\alpha\beta})_{bf}$, $c^{TT}_{\alpha\beta}=10^{-2}(c^{TT}_{\alpha\beta})_{bf}$.  $(a^{T}_{\alpha\beta})_{bf}$, $(c^{TT}_{\alpha\beta})_{bf}$  are the  best fit values listed in Tab.\ref{tab:1}.
 Black line: $\overline{P^{s}}_{\alpha\beta}$ from the best fit values of the global fit data with NO\cite{36}.
 }
\end{figure}

The decoupling pattern can be obtained from the Hamiltonian $H_{LV}$ dominated by the parameter $(H_{LV})_{e\tau}$, namely $|H_{LV}|_{e\tau} \gg |H_{LV}|_{\alpha\beta}$ with $\alpha(\beta)\neq e, \tau$.
According to the expression of $H_{LV}$ shown in Eq.\ref{eq:2}, we know that the texture of $H_{LV}$ is dependent on the neutrino energy $E_{\nu}$.
For the antineutrinos (neutrinos) with $E_{\nu}\geq1\rm TeV$, the decoupling texture can be realised with the LV parameters in the allowed ranges in Tab.\ref{tab:1}.
The specific decoupling energy-scale $E_{d}$ is determined by the  magnitudes of the LV parameters.
Here we consider two representative decoupling energy-scales for antineutrinos (neutrinos). For the LV parameters $a^{T}_{\alpha\beta}=10^{-3}(a^{T}_{\alpha\beta})_{bf}$, $C^{TT}_{\alpha\beta}=10^{-2}(c^{TT}_{\alpha\beta})_{bf}$,
we have $E_{d1}\sim5\rm TeV$. For the parameters  $a^{T}_{\alpha\beta}=10^{-7}(a^{T}_{\alpha\beta})_{bf}$, $C^{TT}_{\alpha\beta}=10^{-7}(c^{TT}_{\alpha\beta})_{bf}$, we obtain $E_{d2}\sim1\rm PeV$.
The dependence of $\overline{P}_{\alpha\beta}$ on $E_{\nu}$ is shown in Fig.\ref{fig:1}.
In the first case, the behavior of $\overline{P}_{\alpha\beta}$ is simple. When $E_{\nu}> E_{d1}$, $\overline{P}_{\alpha\beta}$ shows the decoupling pattern. In the energy range [1TeV, 5TeV],
a small variation is added to the  pattern.
In the second case, the energy $E_{\nu}$ can be classified into 3 areas.
In the range [1TeV, 10TeV], $H_{0}$ is dominant in $H$, namely  $\overline{P}_{\alpha\beta} \sim\overline{P^{s}}_{\alpha\beta}$.
In the range [10TeV, 1PeV], the impact of $H_{LV}$ is noticeable. The variation of  $\overline{P}_{\alpha\beta}$ with $E_{\nu}$ is steep, especially in the range [100TeV, 1PeV].
In the range $E_{\nu}>E_{d2}$, $\overline{P}_{\alpha\beta}$ is also converged to the decoupling pattern.
General speaking, we can see that tiny LV parameters may bring an interesting phenomenology of the flavor transition of HANs, which hence impacts the flux of $\bar{\nu}_{e}$.

\begin{table}
\caption{\label{tab:1} Constraints on the Lorentz violation parameters \cite{33}.}
  \centering
  \begin{tabular}{c c c}
     \noalign{\smallskip}\hline
     \noalign{\smallskip}\hline
     LV Parameters ~~&~~ $95\%$ confidence limit  ~~&~~ Best fit \\
     \noalign{\smallskip}\hline
     Re$(a^{T}_{e\mu})$ ~~~~&~~~~ $1.8\times10^{-23}$ GeV ~~~~&~~~~ $1.0\times10^{-23}$ GeV \\
    Im$(a^{T}_{e\mu})$  ~~~~&~~~~$1.8\times10^{-23}$ GeV ~~~~&~~~~ $4.6\times10^{-24}$ GeV \\
    Re$(c^{TT}_{e\mu})$ ~~~~&~~~~ $8.0\times10^{-27}$ ~~~~&~~~~$1.0\times10^{-28}$ \\
   Im$(c^{TT}_{e\mu})$   ~~~~&~~~~$8.0\times10^{-27}$ ~~~~&~~~~$1.0\times10^{-28}$  \\
   \noalign{\smallskip}\hline
      Re$(a^{T}_{e\tau})$  ~~~~&~~~~$4.1\times10^{-23}$ GeV ~~~~&~~~~ $2.2\times10^{-24}$ GeV\\
     Im$(a^{T}_{e\tau})$~~~~&~~~~ $2.8\times10^{-23}$ GeV ~~~~&~~~~ $1.0\times10^{-28}$ GeV \\
     Re$(c^{TT}_{e\tau})$~~~~&~~~~$9.3\times10^{-25}$ ~~~~&~~~~$1.0\times10^{-28}$  \\
    Im$(c^{TT}_{e\tau})$  ~~~~&~~~~ $1.0\times10^{-24}$ ~~~~&~~~~$3.5\times10^{-25}$  \\
    \noalign{\smallskip}\hline
       Re$(a^{T}_{\mu\tau})$~~~~&~~~~ $6.5\times10^{-24}$ GeV ~~~~&~~~~ $3.2\times10^{-24}$ GeV \\
      Im$(a^{T}_{\mu\tau})$ ~~~~&~~~~$5.1\times10^{-24}$ GeV ~~~~&~~~~ $1.0\times10^{-28}$ GeV \\
       Re$(c^{TT}_{\mu\tau})$ ~~~~&~~~~ ~$4.4\times10^{-27}$ ~~~~&~~~~$1.0\times10^{-28}$ \\
     Im$(c^{TT}_{\mu\tau})$   ~~~~&~~~~ ~$4.2\times10^{-27}$ ~~~~&~~~~$7.5\times10^{-28}$ \\
     \hline
   \end{tabular}
\end{table}

\subsection{The energy spectrum of $\bar{\nu}_{e}$ modified by the LV}
As has been shown, the LV resulting in the $\nu_{\mu}$ ($\bar{\nu}_{\mu}$) decoupling  has significant impacts on the flavor transition of HANs.  Thus, the energy spectrum of $\bar{\nu}_{e}$
may be modified by the LV effect. In addition to $H_{LV}$, the flux of $\bar{\nu}_{e}$ at Earth is also dependent on the production source of HANS.
Since the source of $\bar{\nu}_{e}$ is uncertain at present, here we consider two typical sources to set the original flavor composition.
They are listed as follows: the pion ($\pi^{\pm}$)  decay source with the original flavor ratio
\begin{equation}
\label{eq:16}
R^{s(\pi^{\pm})}=( 1/3, ~2/3,~ 0 ),
\end{equation}
and the muon ($\mu^{\pm}$) damped source with the ratio
\begin{equation}
\label{eq:17}
R^{s(\mu^{\pm})}=( 0,~ 1,~ 0 ).
\end{equation}
For the both sources, the flux of neutrinos is equal to that of antineutrinos.
Employing the flavor ratio at the source and the transition matrix $\overline{P}$, we can derive the ratio of $\bar{\nu}_{e}$ at Earth, namely
\begin{equation}
\label{eq:18}
r_{e}(E_{\nu})=\frac{\phi_{\bar{\nu}_{e}}(E_{\nu})}{\phi_{\nu+\bar{\nu}}(E_{\nu})}=\frac{1}{2}\sum_{\alpha=e,\mu,\tau}\overline{P}_{e\alpha}(E_{\nu})R^{s}_{\alpha},
\end{equation}
in which $\phi_{\bar{\nu}_{e}}$ and $\phi_{\nu+\bar{\nu}}$ are the  differentia flux  of $\bar{\nu}_{e}$ and that of total HANs respectively,
$R^{s}_{\alpha}$ is the component of the flavor ratio vector at the source.
Accordingly, the differentia flux of $\bar{\nu}_{e}$ is predicted to be
\begin{equation}
\label{eq:19}
\phi_{\bar{\nu}_{e}}=\frac{1}{2}\phi_{\nu+\bar{\nu}}(E_{\nu})\times(\sum_{\alpha=e,\mu,\tau}\overline{P}_{e\alpha}(E_{\nu})R^{s}_{\alpha}~).
\end{equation}

At present, the uncertainty of $\phi_{\nu+\bar{\nu}}$ is large. Several models of the energy spectra  can work, e.g., the single power-law,
double power-law, single power law with spectral cutoff, log-parabola, segmented power-law.
For the sake of illustration, we consider the single power-law and the double power-law spectrum in this paper.
They are expressed respectively as\cite{38}
\begin{equation}
\label{eq:20}
\phi_{\nu+\bar{\nu}}^{spl}=\phi_{astro}\times(\frac{E_{\nu}}{100\rm TeV})^{-\gamma_{astro}}\cdot10^{-18}\rm GeV^{-1}cm^{-2}s^{-1}sr^{-1},
\end{equation}
\begin{equation}
\label{eq:21}
\phi_{\nu+\bar{\nu}}^{dpl}=(\phi_{hard}\times(\frac{E_{\nu}}{100\rm TeV})^{-\gamma_{hard}}+\phi_{soft}\times(\frac{E_{\nu}}{100\rm TeV})^{-\gamma_{soft}})\cdot10^{-18}\rm GeV^{-1}cm^{-2}s^{-1}sr^{-1}.
\end{equation}
The ranges of the normalization and the spectral index at the 68.3\% confidence level are listed in Tab.\ref{tab:2}.
\begin{table}
\caption{\label{tab:2} Energy-spectrum parameters of HANs at the 68.3\% confidence level\cite{38}. }
  \centering
  \begin{tabular}{c c c c c }
     \noalign{\smallskip}\hline
     \noalign{\smallskip}\hline
     Model~~&~~~~~~ normalization  ~~&~~~~&~ ~ spectral index  ~&~~ \\
     \noalign{\smallskip}\hline
     single power-law ~~~& $\phi_{astro}$ ~~~&~&~~~ $\gamma_{astro}$ ~ \\
  ~~~~~&$5.68^{+1.56}_{-1.55}$ ~~~&~&~~~~ $2.89^{+0.23}_{-0.20}$ \\
   \noalign{\smallskip}\hline
       double power-law ~~~&$\phi_{hard}$ ~~&~ $\phi_{soft}$ ~~&~~ $\gamma_{hard}$ &~$\gamma_{soft}$~ \\
    ~~~~&$3.26^{+1.88}_{-2.44}$ ~~&~ $0.08^{+3.64}_{-0.08}$~&~~~~~ $2.78^{+0.24}_{-0.32}$~~&~~~~ $3.12^{+1.01}_{-0.30}$\\
     \hline
   \end{tabular}
\end{table}
Using these parameters, we show the LV effect on the energy spectrum of $\bar{\nu}_{e}$ in Fig.\ref{fig:2} and Fig.\ref{fig:3}.

For the  muon ($\mu^{\pm}$) damped source (see Fig.\ref{fig:3}), the LV can bring a noticeable damping of the flux of $\bar{\nu}_{e}$ when $E_{\nu}\geq E_{d}$.
The damping rate $\phi_{\bar{\nu}_{e}}^{LV}$/ $\phi_{\bar{\nu}_{e}}^{S }$ is of order [$10^{-6}$, $10^{-4}$] at the resonance energy. Here $\phi_{\bar{\nu}_{e}}^{LV}$ and $\phi_{\bar{\nu}_{e}}^{S }$ denote the flux including the LV and that without the LV respectively. The specific  damping value is dependent on the form of $\phi_{\nu+\bar{\nu}}$.
For the  pion ($\pi^{\pm}$) decay source (see Fig.\ref{fig:2}), the damping effect is moderate and concealed by the large uncertainty of $\phi_{\nu+\bar{\nu}}$, irrespective of the magnitude of the decoupling energy.
In general, we can expect that  the LV effect can lead to a noticeable decrease of  the $\bar{\nu}_{e}$ flux around the resonance energy for a source dominated by $\bar{\nu}_{\mu}$ ($\nu_{\mu}$).
Thus, the detection of GR events may set stringent constraints on the texture of $H_{LV}$  and the original flavor ratio.

\begin{figure}
\label{fig:2}
  \centering
  \includegraphics[width=.49\textwidth]{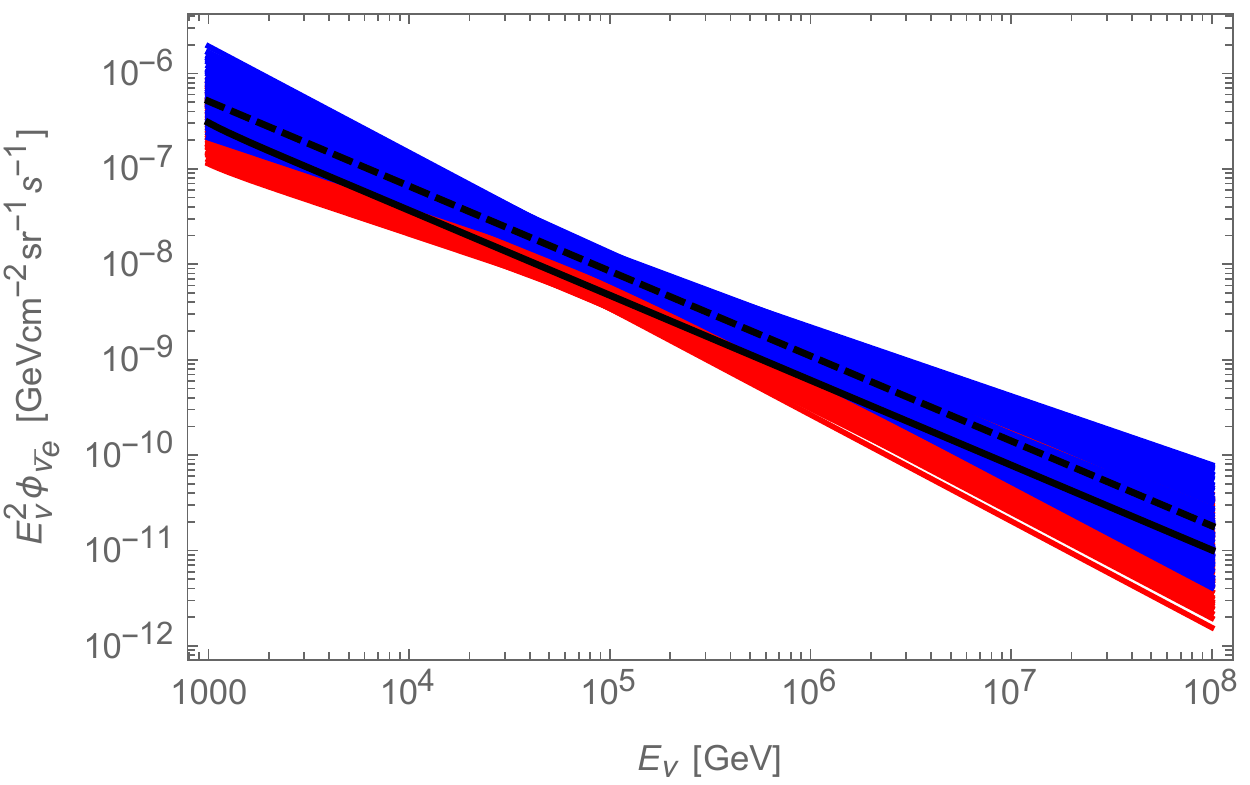}
   \hfill
  \includegraphics[width=.49\textwidth]{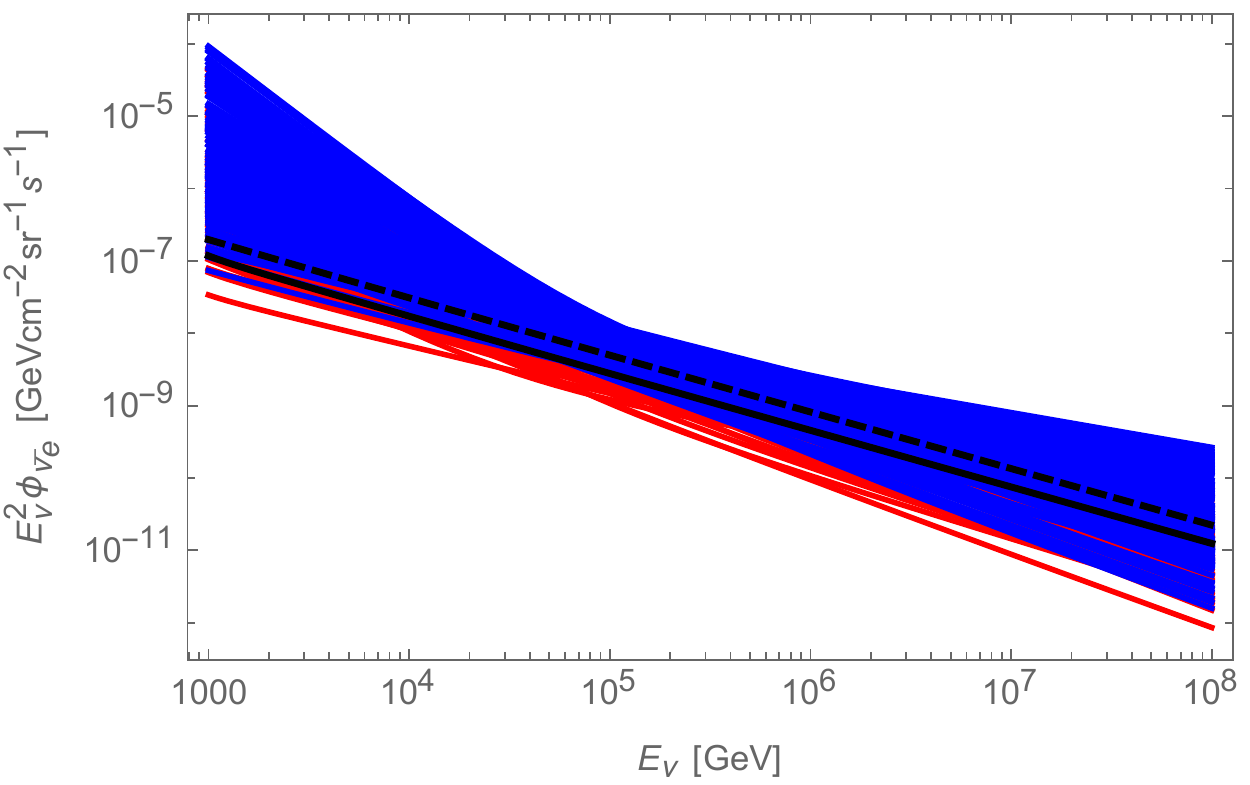}
   \includegraphics[width=.49\textwidth]{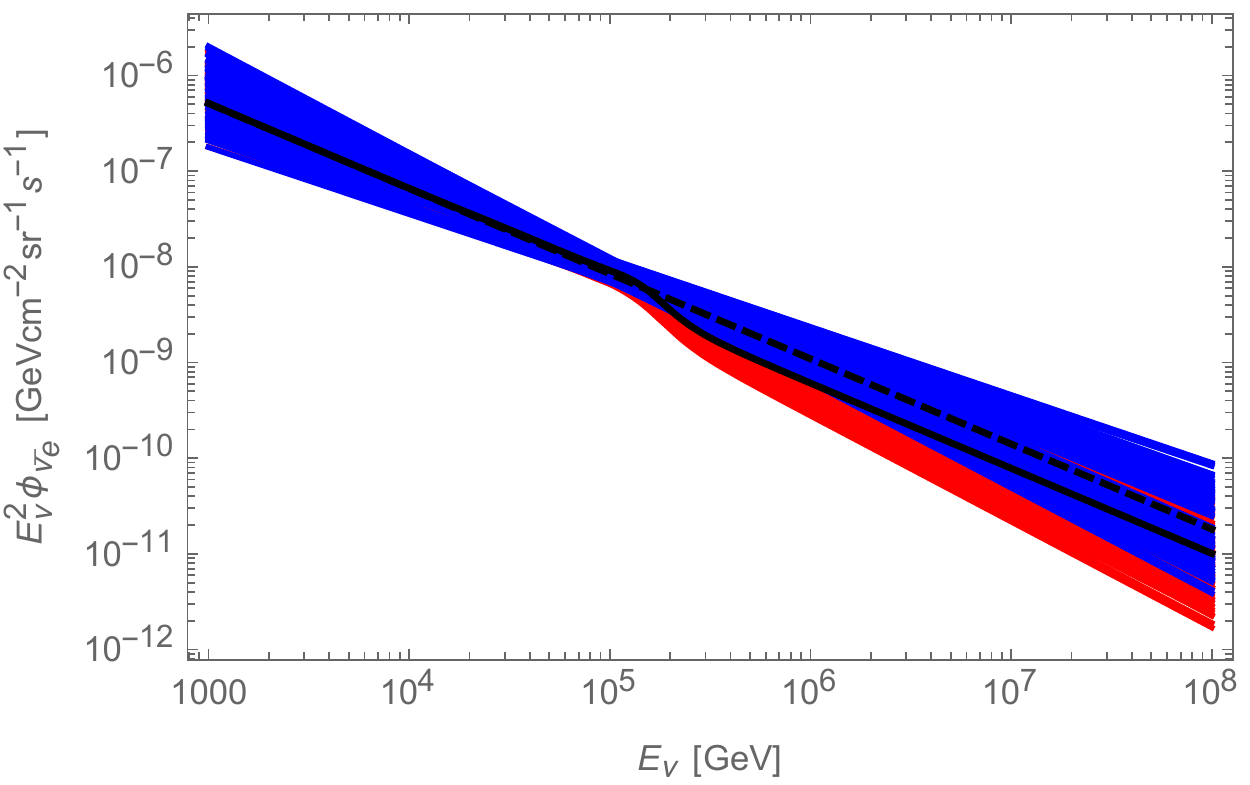}
   \hfill
  \includegraphics[width=.49\textwidth]{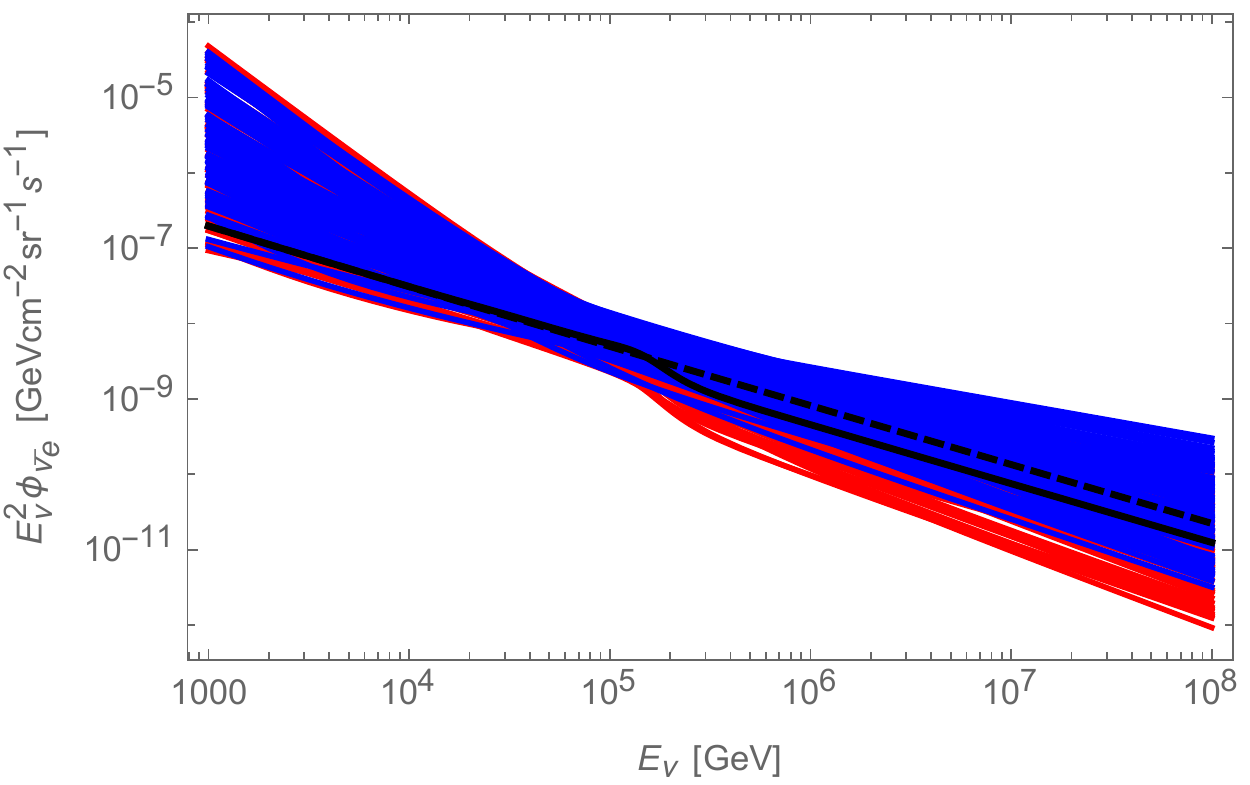}
  \caption{\label{fig:2} The energy spectrum of $\bar{\nu}_{e}$ from the pion ($\pi^{\pm}$) decay source. Left plots: from the total flux of the single power-law. Right plots: from the total flux of the double power-law.
   The ranges of the parameters of the energy spectrum $\phi_{\nu+\bar{\nu}}$ are taken as those listed in Tab.\ref{tab:2}.
   Blue lines: arising from $\overline{P^{s}}$ without the LV effects.  Red lines:  from $\overline{P}$ with  the LV effects. For $\overline{P^{s}}$ and $\overline{P}$,
   the leptonic mixing parameters are in the $3\sigma$ allowed ranges of the global fit
   data with NO\cite{36}. The plots in the first row: with the LV parameters  taken as $a^{T}_{\alpha\beta}=10^{-3}(a^{T}_{\alpha\beta})_{bf}$, $c^{TT}_{\alpha\beta}=10^{-2}(c^{TT}_{\alpha\beta})_{bf}$.
   The plots in the second row: $a^{T}_{\alpha\beta}=10^{-7}(a^{T}_{\alpha\beta})_{bf}$, $c^{TT}_{\alpha\beta}=10^{-7}(c^{TT}_{\alpha\beta})_{bf}$.
  The black line: arising from $\overline{P}$ with the best fit values of the energy spectrum $\phi_{\nu+\bar{\nu}}$ and leptonic mixing parameters. The dashed black line: from $\overline{P^{s}}$ with the best fit values of the leptonic mixing parameters.}
\end{figure}
\begin{figure}
\label{fig:3}
  \centering
  \includegraphics[width=.49\textwidth]{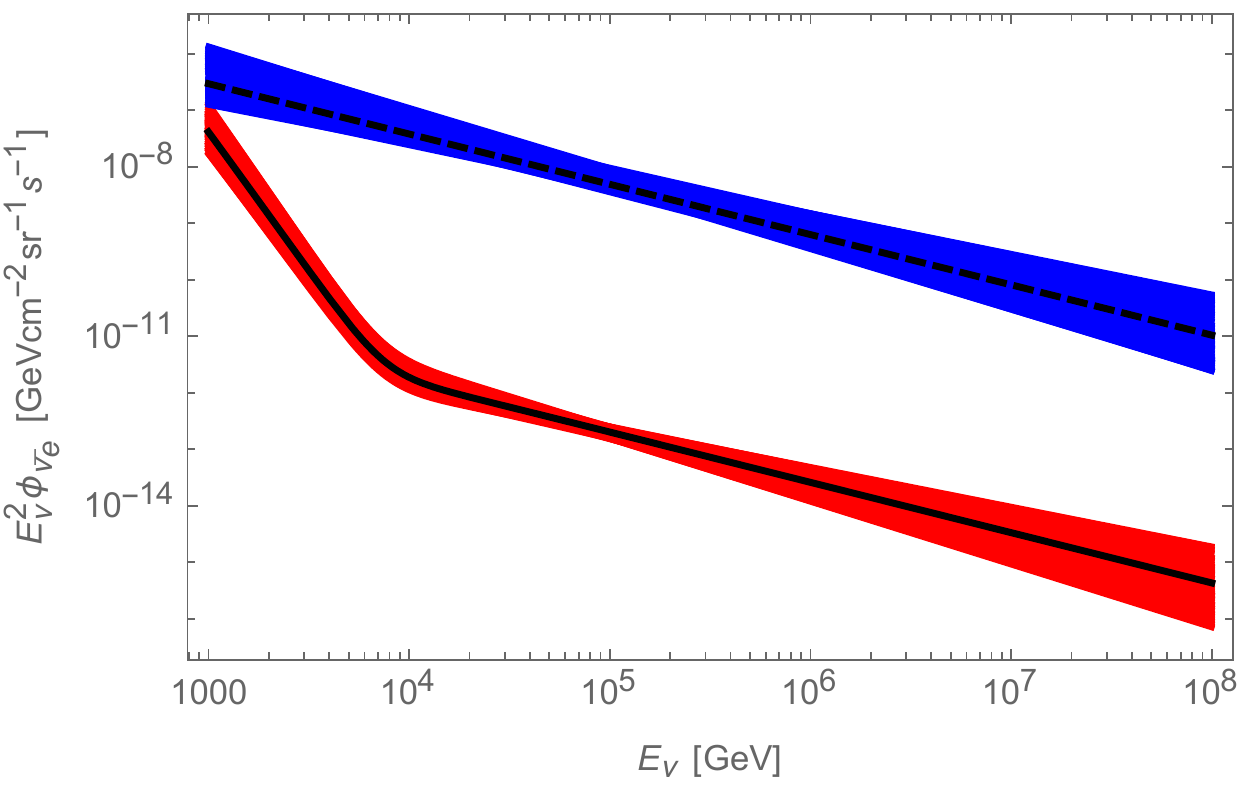}
   \hfill
  \includegraphics[width=.49\textwidth]{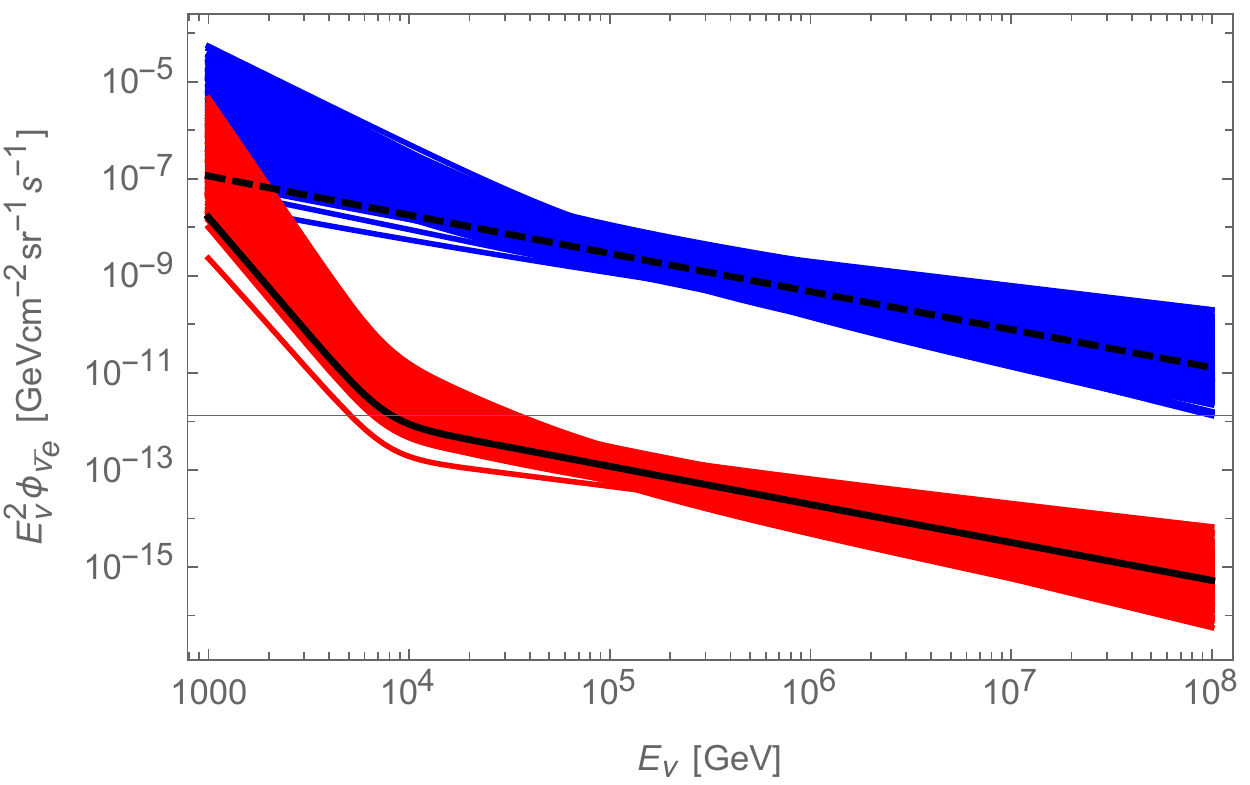}
  \includegraphics[width=.49\textwidth]{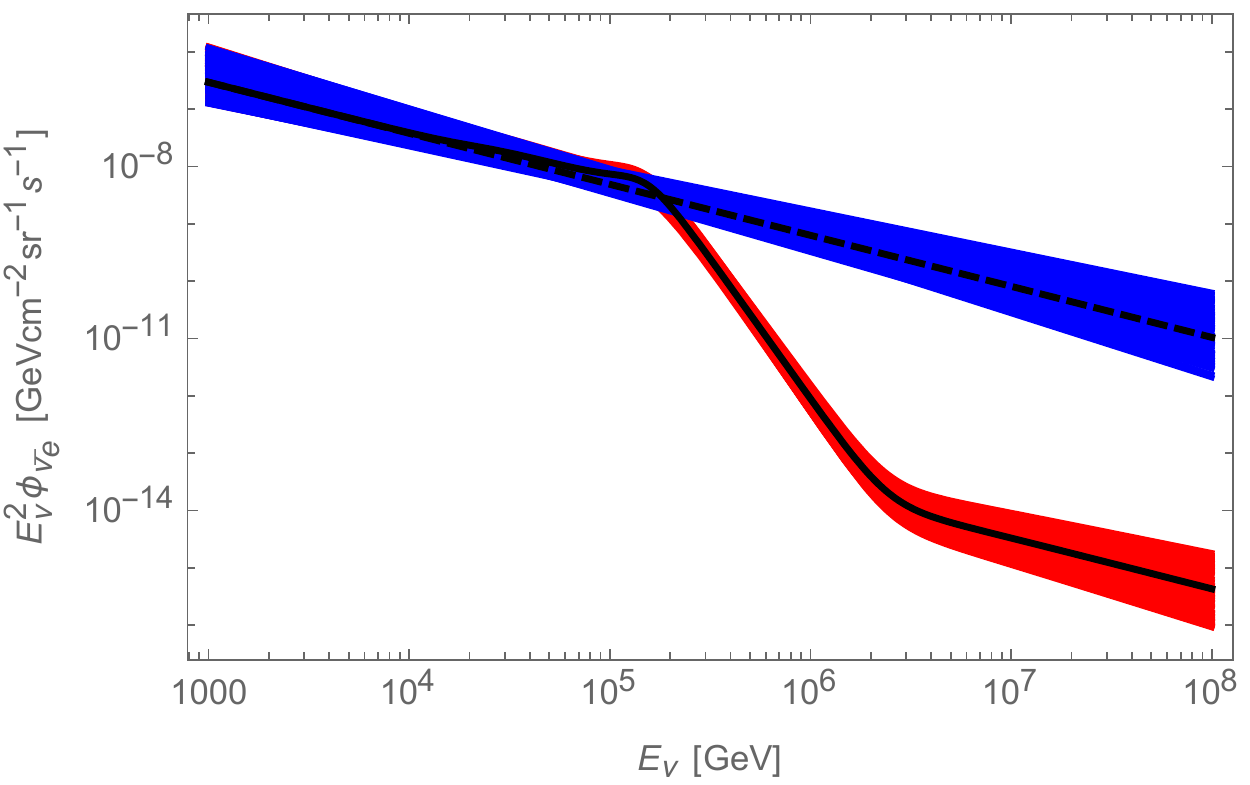}
   \hfill
  \includegraphics[width=.49\textwidth]{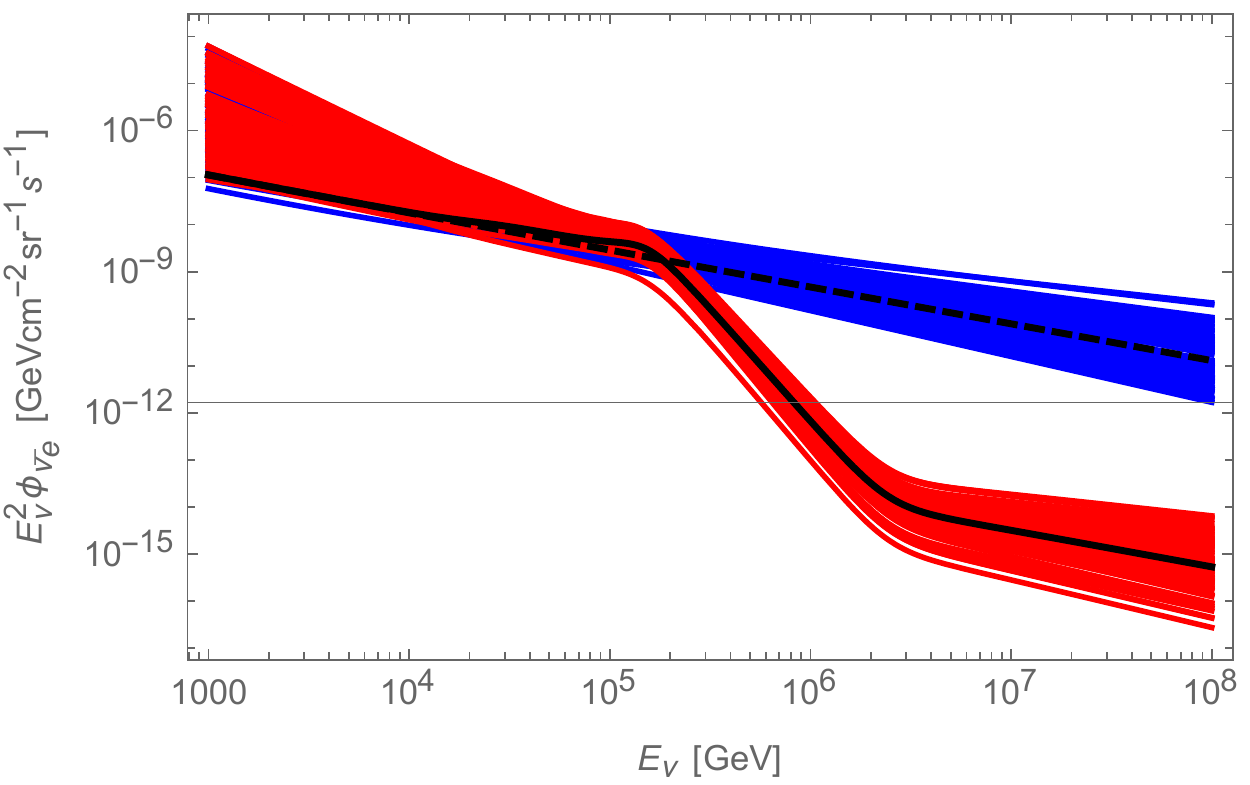}
  \caption{\label{fig:3} The energy spectrum of $\bar{\nu}_{e}$ from the muon ($\mu^{\pm}$) damped source. The conventions of the parameters, colors and the lines are the same as those in Fig.\ref{fig:2}.}
\end{figure}

\subsection{The strength of the GR event rate damped by the LV}
The GR events provide a window to explore the texture of $H_{LV}$ and the flavor composition of the HANs at the PeV energy-scale.
Now let us examine the impacts of the decoupling of $\nu_{\mu}$ ($\bar{\nu}_{\mu}$) on the GR event rate.
To weaken the influence of the uncertain energy spectrum  $\phi_{\nu+\bar{\nu}}$,  for down-coming events, the ratio of the resonance events to the nonresonant continuum events is proposed
to describe the strength of the GR event rate, i.e.\cite{2},
\begin{equation}
\label{eq:22}
\begin{aligned}
  \frac{N_{Res}}{N_{non-Res}(E_{\nu}>E_{\nu}^{min})}&=\frac{10\pi}{18}(\frac{\Gamma_{W}}{M_{W}})(\frac{\sigma^{peak}_{Res}}{\sigma^{CC}_{\nu N}(E_{\nu}=6.3\rm PeV)})\frac{(\alpha-1.4)(\frac{E_{\nu}^{min}}{6.3\rm PeV})^{\alpha-1.4}}{[1-(\frac{E_{\nu}^{min}}{E_{\nu}^{max}})^{(\alpha-1.4)}]}\times [r_{e}]_{E_{\nu}=6.3\rm PeV} \\
  &=11\times\frac{(\alpha-1.4)(\frac{E_{\nu}^{min}}{6.3\rm PeV})^{\alpha-1.4}}{[1-(\frac{E_{\nu}^{min}}{E_{\nu}^{max}})^{(\alpha-1.4)}]}\times [r_{e}]_{E_{\nu}=6.3\rm PeV}.
\end{aligned}
\end{equation}
Here the spectral index $\alpha$ is dependent on the acceleration mechanism of neutrinos. In absence of GR events, $\alpha\geq2.3$\cite{39,40}. In this paper, we take $\alpha=2$.
We note that a moderate variation of $\alpha$ will not lead to a noticeable change of the observation on the LV effect.
Employing the best fit values of the leptonic mixing parameters with NO\cite{36} and the LV parameters $a^{T}_{\alpha\beta}=10^{-3}(a^{T}_{\alpha\beta})_{bf}$, $c^{TT}_{\alpha\beta}=10^{-2}(c^{TT}_{\alpha\beta})_{bf}$ with $(a^{T}_{\alpha\beta})_{bf}$, $(c^{TT}_{\alpha\beta})_{bf}$ listed in Tab.\ref{tab:1}, we calculate the ratio of resonance event rate around the resonance energy.
The specific values are listed in Tab.\ref{tab:3}. For the sake of comparison, the parenthetic values obtained from the standard flavor transition matrix $\overline{P^{s}}$ are also shown.
\begin{table}
\caption{\label{tab:3} The ratio of resonance event rate to nonresonant event rate with and without the LV effect for the  pion ($\pi^{\pm}$) decay source, muon ($\mu^{\pm}$) damped source, and a source rich in $\nu_{\mu}$ ($\bar{\nu}_{\mu}$) with $\phi_{\nu}=\phi_{\bar{\nu}}$. Here $\alpha=2$, $E_{\nu}^{min}$=1, 2, 3, 4, 5 PeV, $E_{\nu}^{max}=\infty$, $r_{e}$ takes the value $ [r_{e}]_{E_{\nu}=6.3\rm PeV}$.
The parenthetic values are obtained from the standard flavor transition matrix $\overline{P^{s}}$. }
  \centering
  \begin{tabular}{c c c c c c c}
     \noalign{\smallskip}\hline
     \noalign{\smallskip}\hline
     $E_{\nu}^{min}(\rm PeV)$ ~&~~1  ~&~2~~&~~3~~&~~ 4~~&~~ 5&~~ \\
     \noalign{\smallskip}\hline
$R^{s(\pi^{\pm})}=( 1/3, ~2/3,~ 0 )$ ~&~~0.18~~&~~~0.28~ &~~~0.35 &~~~0.42&~~~0.48&~$r_{e}$=0.083\\
~~&~~(0.32)~~&~~~(0.50)~ &~~~(0.63) &~~~(0.75)&~~~(0.86)&~($r_{e}$=0.149)\\ \hline
$R^{s(\mu^{\pm})}=( 0,~ 1,~ 0 )$  ~~~~~&~~$7.8\times10^{-6}$~~~~&~~$1.2\times10^{-5}$~~&~$1.5\times10^{-5}$&~$1.8\times10^{-5}$ &~$2.0\times10^{-5}$&~$r_{e}=3.5\times10^{-6}$\\
~~~~&~~~~(0.19)~~~~&~~~(0.28)~~&~(0.36)&~(0.43) &~(0.49)&~($r_{e}$=0.086)\\ \hline
$R^{s(x)}=( 0.1,~ 0.9,~ 0 )$ ~~&~~0.05~~&~~~0.08~~ &~~0.10~~&~~0.12~~&~~0.14~~&~$r_{e}$=0.025\\
~~~~&~~~~~(0.23)~~~~&~~~(0.35)~~&~~(0.44)&~~(0.53) &~~(0.60)&~~($r_{e}$=0.105)\\
     \hline
   \end{tabular}
\end{table}

From Tab.\ref{tab:3}, we can see that the ratio of  resonance event rate to the nonresonant event rate
is mainly determined by the ratio of $\bar{\nu}_{e}$ at the resonance energy-scale. The damping rate of the GR event rate approximates that of $ [r_{e}]_{E_{\nu}=6.3\rm PeV}$,
namely
\begin{equation}
\label{eq:23}
  (\frac{N_{Res}}{N_{non-Res}(E_{\nu}>E_{\nu}^{min})})^{LV}/(\frac{N_{Res}}{N_{non-Res}(E_{\nu}>E_{\nu}^{min})})^{S}\sim ([r_{e}]_{E_{\nu}=6.3\rm PeV})^{LV}/([r_{e}]_{E_{\nu}=6.3\rm PeV})^{S},
\end{equation}
in which the superscript 'LV' ('S') denotes the value with (without) the LV effect.
For the original flavor ratios $R^{s(\pi^{\pm})}$, $R^{s(\mu^{\pm})}$, and $R^{s(x)}$, $([r_{e}]_{E_{\nu}=6.3\rm PeV})^{LV}/([r_{e}]_{E_{\nu}=6.3\rm PeV})^{S}$ takes 0.557, $4.07\times10^{-5}$, 0.238, respectively.
The damping rate is sensitive to the increase of the original ratio of $\bar{\nu}_{\mu}$ around the resonance energy when $\phi_{\bar{\nu}_{\mu}}/\phi_{\bar{\nu}}>0.9$.
Furthermore, when the decoupling energy is around $1\rm PeV$, i.e., $a^{T}_{\alpha\beta}=10^{-7}(a^{T}_{\alpha\beta})_{bf}$, $c^{TT}_{\alpha\beta}=10^{-7}(c^{TT}_{\alpha\beta})_{bf}$,
the values of  $([r_{e}]_{E_{\nu}=6.3\rm PeV})^{LV}$ are similar to those listed in Tab.\ref{tab:3}. Thus, the damping rate of the GR event rate is mainly
sensitive to the texture of $H_{LV}$, irrespective the magnitudes of the LV parameters.
However, we should keep in mind  that the magnitudes of the  parameters determine the decoupling energy which impacts the evolution of $\overline{P}$ with $E_{\nu}$.
In the range $E_{\nu}<E_{d}$, the flavor ratio of HANs at Earth is dependent on the magnitudes of the LV parameters.

\subsection{Comments on the realistic source and the $\nu_{\tau}$ event at the PeV energy-scale }

At present, the production source and the acceleration mechanism of HANs are unknown. However, it is widely believed that the strong magnetic field plays an important role in the
acceleration of HANs\cite{41}, e.g. acceleration by the magnetic reconnection\cite{42,43,44}. As is known, the synchrotron radiation of the charged particle in the strong magnetic field becomes significant at high energies. \
We can expect that the muon ($\mu^{\pm}$) damped source may be dominant at the PeV energy-scale.  Accordingly, when $E_{d}<6.3\rm PeV$, the decoupling pattern can show a noticeable impact on the GR events.
Furthermore, we note that the decoupling pattern can bring the following observation:
\begin{equation}
\label{eq:24}
\phi_{\nu_{e}}\sim\phi_{\nu_{\tau}},~~~~\phi_{\bar{\nu}_{e}}\sim\phi_{\bar{\nu}_{\tau}}.
\end{equation}
Hence, for the muon ($\mu^{\pm}$) damped source, the proposed LV texture can lead to a notable decrease of both the GR event rate and the double bang events from $\nu_{\tau}$ at the PeV energy-scale.
In other words, we can obtain the double constraints on the texture of $H_{LV}$ from the GR event and the double bang event.

\section{Conclusions }

The Lorentz invariance is a fundamental symmetry in nature. Strong constraints on the LV parameters have been obtained at low energies.
To connect the LV to the GR event at the PeV energy-scale, we proposed the LV Hamiltonian of the special texture which can  bring the $\nu_{\mu}$ ($\bar{\nu}_{\mu}$) decoupling  at high energies.
On the base of the $\nu_{\mu}$ ($\bar{\nu}_{\mu}$) decoupling, we analysed the impacts of the LV parameters on the flavor transition of HANs.
For the typical sources of HANs, the damping effect from the decoupling pattern on the energy spectrum of $\bar{\nu}_{e}$ was shown.
For the  muon ($\mu^{\pm}$) damped source, the decrease of the $\bar{\nu}_{e}$ flux at the PeV energy-scale is notable, irrespective of the form of the total flux $\phi_{\nu+\bar{\nu}}$.
The LV effect can lead to a noticeable damping of the GR event rate if the HANs are mainly from the source dominated by $\bar{\nu}_{\mu}$ ($\nu_{\mu}$).
Furthermore, the decoupling pattern can also decrease the double bang events from $\nu_{\tau}$ at the PeV energy-scale.
Therefore, the detection of GR events and other high-energy events may bring extra constraints on the LV Hamiltonian and the production source of HANs.

\acknowledgments
 This work is supported by the National Natural Science Foundation of China under grant No. 12065007, the Guangxi Scientific Programm Foundation under grant No. Guike AD19110045.
\\

\bibliography{refsa}

\begin{thebibliography}{44}%
\makeatletter
\providecommand \@ifxundefined [1]{%
 \@ifx{#1\undefined}
}%
\providecommand \@ifnum [1]{%
 \ifnum #1\expandafter \@firstoftwo
 \else \expandafter \@secondoftwo
 \fi
}%
\providecommand \@ifx [1]{%
 \ifx #1\expandafter \@firstoftwo
 \else \expandafter \@secondoftwo
 \fi
}%
\providecommand \natexlab [1]{#1}%
\providecommand \enquote  [1]{``#1''}%
\providecommand \bibnamefont  [1]{#1}%
\providecommand \bibfnamefont [1]{#1}%
\providecommand \citenamefont [1]{#1}%
\providecommand \href@noop [0]{\@secondoftwo}%
\providecommand \href [0]{\begingroup \@sanitize@url \@href}%
\providecommand \@href[1]{\@@startlink{#1}\@@href}%
\providecommand \@@href[1]{\endgroup#1\@@endlink}%
\providecommand \@sanitize@url [0]{\catcode `\\12\catcode `\$12\catcode
  `\&12\catcode `\#12\catcode `\^12\catcode `\_12\catcode `\%12\relax}%
\providecommand \@@startlink[1]{}%
\providecommand \@@endlink[0]{}%
\providecommand \url  [0]{\begingroup\@sanitize@url \@url }%
\providecommand \@url [1]{\endgroup\@href {#1}{\urlprefix }}%
\providecommand \urlprefix  [0]{URL }%
\providecommand \Eprint [0]{\href }%
\providecommand \doibase [0]{http://dx.doi.org/}%
\providecommand \selectlanguage [0]{\@gobble}%
\providecommand \bibinfo  [0]{\@secondoftwo}%
\providecommand \bibfield  [0]{\@secondoftwo}%
\providecommand \translation [1]{[#1]}%
\providecommand \BibitemOpen [0]{}%
\providecommand \bibitemStop [0]{}%
\providecommand \bibitemNoStop [0]{.\EOS\space}%
\providecommand \EOS [0]{\spacefactor3000\relax}%
\providecommand \BibitemShut  [1]{\csname bibitem#1\endcsname}%
\let\auto@bib@innerbib\@empty
\bibitem [{\citenamefont {Glashow}(1960)}]{1}%
  \BibitemOpen
  \bibfield  {author} {\bibinfo {author} {\bibfnamefont {S.~L.}\ \bibnamefont
  {Glashow}},\ }\href@noop {} {\bibfield  {journal} {\bibinfo  {journal} {Phys.
  Rev.}\ }\textbf {\bibinfo {volume} {118}},\ \bibinfo {pages} {316} (\bibinfo
  {year} {1960})}\BibitemShut {NoStop}%
\bibitem [{\citenamefont {Barger}\ \emph {et~al.}(2014)\citenamefont {Barger},
  \citenamefont {Fu}, \citenamefont {Learned}, \citenamefont {Marfatia},
  \citenamefont {Pakvasa},\ and\ \citenamefont {Weiler}}]{2}%
  \BibitemOpen
  \bibfield  {author} {\bibinfo {author} {\bibfnamefont {V.}~\bibnamefont
  {Barger}}, \bibinfo {author} {\bibfnamefont {L.}~\bibnamefont {Fu}}, \bibinfo
  {author} {\bibfnamefont {J.~G.}\ \bibnamefont {Learned}}, \bibinfo {author}
  {\bibfnamefont {D.}~\bibnamefont {Marfatia}}, \bibinfo {author}
  {\bibfnamefont {S.}~\bibnamefont {Pakvasa}}, \ and\ \bibinfo {author}
  {\bibfnamefont {T.~J.}\ \bibnamefont {Weiler}},\ }\href@noop {} {\bibfield
  {journal} {\bibinfo  {journal} {Phys. Rev. D}\ }\textbf {\bibinfo {volume}
  {90}},\ \bibinfo {pages} {121301} (\bibinfo {year} {2014})},\ \Eprint
  {http://arxiv.org/abs/1407.3255} {arXiv:1407.3255 [astro-ph.HE]} \BibitemShut
  {NoStop}%
\bibitem [{\citenamefont {Huang}\ and\ \citenamefont {Liu}(2020)}]{3}%
  \BibitemOpen
  \bibfield  {author} {\bibinfo {author} {\bibfnamefont {G.-y.}\ \bibnamefont
  {Huang}}\ and\ \bibinfo {author} {\bibfnamefont {Q.}~\bibnamefont {Liu}},\
  }\href@noop {} {\bibfield  {journal} {\bibinfo  {journal} {JCAP}\ }\textbf
  {\bibinfo {volume} {03}},\ \bibinfo {pages} {005} (\bibinfo {year} {2020})},\
  \Eprint {http://arxiv.org/abs/1912.02976} {arXiv:1912.02976 [hep-ph]}
  \BibitemShut {NoStop}%
\bibitem [{\citenamefont {Abbasi}\ \emph {et~al.}(2022)\citenamefont {Abbasi}
  \emph {et~al.}}]{4}%
  \BibitemOpen
  \bibfield  {author} {\bibinfo {author} {\bibfnamefont {R.}~\bibnamefont
  {Abbasi}} \emph {et~al.} (\bibinfo {collaboration} {IceCube}),\ }\href@noop
  {} {\bibfield  {journal} {\bibinfo  {journal} {Astrophys. J.}\ }\textbf
  {\bibinfo {volume} {928}},\ \bibinfo {pages} {50} (\bibinfo {year} {2022})},\
  \Eprint {http://arxiv.org/abs/2111.10299} {arXiv:2111.10299 [astro-ph.HE]}
  \BibitemShut {NoStop}%
\bibitem [{\citenamefont {Aartsen}\ \emph {et~al.}(2021)\citenamefont {Aartsen}
  \emph {et~al.}}]{5}%
  \BibitemOpen
  \bibfield  {author} {\bibinfo {author} {\bibfnamefont {M.~G.}\ \bibnamefont
  {Aartsen}} \emph {et~al.} (\bibinfo {collaboration} {IceCube}),\ }\href@noop
  {} {\bibfield  {journal} {\bibinfo  {journal} {Nature}\ }\textbf {\bibinfo
  {volume} {591}},\ \bibinfo {pages} {220} (\bibinfo {year} {2021})},\ \Eprint
  {http://arxiv.org/abs/2110.15051} {arXiv:2110.15051 [hep-ex]} \BibitemShut
  {NoStop}%
\bibitem [{\citenamefont {Kobayashi}\ and\ \citenamefont {Lim}(2001)}]{6}%
  \BibitemOpen
  \bibfield  {author} {\bibinfo {author} {\bibfnamefont {M.}~\bibnamefont
  {Kobayashi}}\ and\ \bibinfo {author} {\bibfnamefont {C.~S.}\ \bibnamefont
  {Lim}},\ }\href@noop {} {\bibfield  {journal} {\bibinfo  {journal} {Phys.
  Rev. D}\ }\textbf {\bibinfo {volume} {64}},\ \bibinfo {pages} {013003}
  (\bibinfo {year} {2001})},\ \Eprint {http://arxiv.org/abs/hep-ph/0012266}
  {arXiv:hep-ph/0012266} \BibitemShut {NoStop}%
\bibitem [{\citenamefont {Beacom}\ \emph {et~al.}(2004)\citenamefont {Beacom},
  \citenamefont {Bell}, \citenamefont {Hooper}, \citenamefont {Learned},
  \citenamefont {Pakvasa},\ and\ \citenamefont {Weiler}}]{7}%
  \BibitemOpen
  \bibfield  {author} {\bibinfo {author} {\bibfnamefont {J.~F.}\ \bibnamefont
  {Beacom}}, \bibinfo {author} {\bibfnamefont {N.~F.}\ \bibnamefont {Bell}},
  \bibinfo {author} {\bibfnamefont {D.}~\bibnamefont {Hooper}}, \bibinfo
  {author} {\bibfnamefont {J.~G.}\ \bibnamefont {Learned}}, \bibinfo {author}
  {\bibfnamefont {S.}~\bibnamefont {Pakvasa}}, \ and\ \bibinfo {author}
  {\bibfnamefont {T.~J.}\ \bibnamefont {Weiler}},\ }\href@noop {} {\bibfield
  {journal} {\bibinfo  {journal} {Phys. Rev. Lett.}\ }\textbf {\bibinfo
  {volume} {92}},\ \bibinfo {pages} {011101} (\bibinfo {year} {2004})},\
  \Eprint {http://arxiv.org/abs/hep-ph/0307151} {arXiv:hep-ph/0307151}
  \BibitemShut {NoStop}%
\bibitem [{\citenamefont {Esmaili}(2010)}]{8}%
  \BibitemOpen
  \bibfield  {author} {\bibinfo {author} {\bibfnamefont {A.}~\bibnamefont
  {Esmaili}},\ }\href@noop {} {\bibfield  {journal} {\bibinfo  {journal} {Phys.
  Rev. D}\ }\textbf {\bibinfo {volume} {81}},\ \bibinfo {pages} {013006}
  (\bibinfo {year} {2010})},\ \Eprint {http://arxiv.org/abs/0909.5410}
  {arXiv:0909.5410 [hep-ph]} \BibitemShut {NoStop}%
\bibitem [{\citenamefont {Brdar}\ and\ \citenamefont {Hansen}(2019)}]{9}%
  \BibitemOpen
  \bibfield  {author} {\bibinfo {author} {\bibfnamefont {V.}~\bibnamefont
  {Brdar}}\ and\ \bibinfo {author} {\bibfnamefont {R.~S.~L.}\ \bibnamefont
  {Hansen}},\ }\href@noop {} {\bibfield  {journal} {\bibinfo  {journal} {JCAP}\
  }\textbf {\bibinfo {volume} {02}},\ \bibinfo {pages} {023} (\bibinfo {year}
  {2019})},\ \Eprint {http://arxiv.org/abs/1812.05541} {arXiv:1812.05541
  [hep-ph]} \BibitemShut {NoStop}%
\bibitem [{\citenamefont {Beacom}\ \emph {et~al.}(2003)\citenamefont {Beacom},
  \citenamefont {Bell}, \citenamefont {Hooper}, \citenamefont {Pakvasa},\ and\
  \citenamefont {Weiler}}]{10}%
  \BibitemOpen
  \bibfield  {author} {\bibinfo {author} {\bibfnamefont {J.~F.}\ \bibnamefont
  {Beacom}}, \bibinfo {author} {\bibfnamefont {N.~F.}\ \bibnamefont {Bell}},
  \bibinfo {author} {\bibfnamefont {D.}~\bibnamefont {Hooper}}, \bibinfo
  {author} {\bibfnamefont {S.}~\bibnamefont {Pakvasa}}, \ and\ \bibinfo
  {author} {\bibfnamefont {T.~J.}\ \bibnamefont {Weiler}},\ }\href@noop {}
  {\bibfield  {journal} {\bibinfo  {journal} {Phys. Rev. Lett.}\ }\textbf
  {\bibinfo {volume} {90}},\ \bibinfo {pages} {181301} (\bibinfo {year}
  {2003})},\ \Eprint {http://arxiv.org/abs/hep-ph/0211305}
  {arXiv:hep-ph/0211305} \BibitemShut {NoStop}%
\bibitem [{\citenamefont {Meloni}\ and\ \citenamefont {Ohlsson}(2007)}]{11}%
  \BibitemOpen
  \bibfield  {author} {\bibinfo {author} {\bibfnamefont {D.}~\bibnamefont
  {Meloni}}\ and\ \bibinfo {author} {\bibfnamefont {T.}~\bibnamefont
  {Ohlsson}},\ }\href {\doibase 10.1103/PhysRevD.75.125017} {\bibfield
  {journal} {\bibinfo  {journal} {Phys. Rev. D}\ }\textbf {\bibinfo {volume}
  {75}},\ \bibinfo {pages} {125017} (\bibinfo {year} {2007})},\ \Eprint
  {http://arxiv.org/abs/hep-ph/0612279} {arXiv:hep-ph/0612279} \BibitemShut
  {NoStop}%
\bibitem [{\citenamefont {Baerwald}\ \emph {et~al.}(2012)\citenamefont
  {Baerwald}, \citenamefont {Bustamante},\ and\ \citenamefont {Winter}}]{12}%
  \BibitemOpen
  \bibfield  {author} {\bibinfo {author} {\bibfnamefont {P.}~\bibnamefont
  {Baerwald}}, \bibinfo {author} {\bibfnamefont {M.}~\bibnamefont
  {Bustamante}}, \ and\ \bibinfo {author} {\bibfnamefont {W.}~\bibnamefont
  {Winter}},\ }\href@noop {} {\bibfield  {journal} {\bibinfo  {journal} {JCAP}\
  }\textbf {\bibinfo {volume} {10}},\ \bibinfo {pages} {020} (\bibinfo {year}
  {2012})},\ \Eprint {http://arxiv.org/abs/1208.4600} {arXiv:1208.4600
  [astro-ph.CO]} \BibitemShut {NoStop}%
\bibitem [{\citenamefont {Pagliaroli}\ \emph {et~al.}(2015)\citenamefont
  {Pagliaroli}, \citenamefont {Palladino}, \citenamefont {Villante},\ and\
  \citenamefont {Vissani}}]{13}%
  \BibitemOpen
  \bibfield  {author} {\bibinfo {author} {\bibfnamefont {G.}~\bibnamefont
  {Pagliaroli}}, \bibinfo {author} {\bibfnamefont {A.}~\bibnamefont
  {Palladino}}, \bibinfo {author} {\bibfnamefont {F.~L.}\ \bibnamefont
  {Villante}}, \ and\ \bibinfo {author} {\bibfnamefont {F.}~\bibnamefont
  {Vissani}},\ }\href@noop {} {\bibfield  {journal} {\bibinfo  {journal} {Phys.
  Rev. D}\ }\textbf {\bibinfo {volume} {92}},\ \bibinfo {pages} {113008}
  (\bibinfo {year} {2015})},\ \Eprint {http://arxiv.org/abs/1506.02624}
  {arXiv:1506.02624 [hep-ph]} \BibitemShut {NoStop}%
\bibitem [{\citenamefont {Bustamante}\ \emph {et~al.}(2017)\citenamefont
  {Bustamante}, \citenamefont {Beacom},\ and\ \citenamefont {Murase}}]{14}%
  \BibitemOpen
  \bibfield  {author} {\bibinfo {author} {\bibfnamefont {M.}~\bibnamefont
  {Bustamante}}, \bibinfo {author} {\bibfnamefont {J.~F.}\ \bibnamefont
  {Beacom}}, \ and\ \bibinfo {author} {\bibfnamefont {K.}~\bibnamefont
  {Murase}},\ }\href@noop {} {\bibfield  {journal} {\bibinfo  {journal} {Phys.
  Rev. D}\ }\textbf {\bibinfo {volume} {95}},\ \bibinfo {pages} {063013}
  (\bibinfo {year} {2017})},\ \Eprint {http://arxiv.org/abs/1610.02096}
  {arXiv:1610.02096 [astro-ph.HE]} \BibitemShut {NoStop}%
\bibitem [{\citenamefont {Denton}\ and\ \citenamefont {Tamborra}(2018)}]{15}%
  \BibitemOpen
  \bibfield  {author} {\bibinfo {author} {\bibfnamefont {P.~B.}\ \bibnamefont
  {Denton}}\ and\ \bibinfo {author} {\bibfnamefont {I.}~\bibnamefont
  {Tamborra}},\ }\href@noop {} {\bibfield  {journal} {\bibinfo  {journal}
  {Phys. Rev. Lett.}\ }\textbf {\bibinfo {volume} {121}},\ \bibinfo {pages}
  {121802} (\bibinfo {year} {2018})},\ \Eprint
  {http://arxiv.org/abs/1805.05950} {arXiv:1805.05950 [hep-ph]} \BibitemShut
  {NoStop}%
\bibitem [{\citenamefont {Ng}\ and\ \citenamefont {Beacom}(2014)}]{16}%
  \BibitemOpen
  \bibfield  {author} {\bibinfo {author} {\bibfnamefont {K.~C.~Y.}\
  \bibnamefont {Ng}}\ and\ \bibinfo {author} {\bibfnamefont {J.~F.}\
  \bibnamefont {Beacom}},\ }\href@noop {} {\bibfield  {journal} {\bibinfo
  {journal} {Phys. Rev. D}\ }\textbf {\bibinfo {volume} {90}},\ \bibinfo
  {pages} {065035} (\bibinfo {year} {2014})},\ \Eprint
  {http://arxiv.org/abs/1404.2288} {arXiv:1404.2288 [astro-ph.HE]} \BibitemShut
  {NoStop}%
\bibitem [{\citenamefont {Ioka}\ and\ \citenamefont {Murase}(2014)}]{17}%
  \BibitemOpen
  \bibfield  {author} {\bibinfo {author} {\bibfnamefont {K.}~\bibnamefont
  {Ioka}}\ and\ \bibinfo {author} {\bibfnamefont {K.}~\bibnamefont {Murase}},\
  }\href@noop {} {\bibfield  {journal} {\bibinfo  {journal} {PTEP}\ }\textbf
  {\bibinfo {volume} {2014}},\ \bibinfo {pages} {061E01} (\bibinfo {year}
  {2014})},\ \Eprint {http://arxiv.org/abs/1404.2279} {arXiv:1404.2279
  [astro-ph.HE]} \BibitemShut {NoStop}%
\bibitem [{\citenamefont {Bustamante}\ \emph {et~al.}(2020)\citenamefont
  {Bustamante}, \citenamefont {Rosenstr\o{}m}, \citenamefont {Shalgar},\ and\
  \citenamefont {Tamborra}}]{18}%
  \BibitemOpen
  \bibfield  {author} {\bibinfo {author} {\bibfnamefont {M.}~\bibnamefont
  {Bustamante}}, \bibinfo {author} {\bibfnamefont {C.}~\bibnamefont
  {Rosenstr\o{}m}}, \bibinfo {author} {\bibfnamefont {S.}~\bibnamefont
  {Shalgar}}, \ and\ \bibinfo {author} {\bibfnamefont {I.}~\bibnamefont
  {Tamborra}},\ }\href@noop {} {\bibfield  {journal} {\bibinfo  {journal}
  {Phys. Rev. D}\ }\textbf {\bibinfo {volume} {101}},\ \bibinfo {pages}
  {123024} (\bibinfo {year} {2020})},\ \Eprint
  {http://arxiv.org/abs/2001.04994} {arXiv:2001.04994 [astro-ph.HE]}
  \BibitemShut {NoStop}%
\bibitem [{\citenamefont {Arg\"uelles}\ \emph {et~al.}(2015)\citenamefont
  {Arg\"uelles}, \citenamefont {Katori},\ and\ \citenamefont {Salvado}}]{19}%
  \BibitemOpen
  \bibfield  {author} {\bibinfo {author} {\bibfnamefont {C.~A.}\ \bibnamefont
  {Arg\"uelles}}, \bibinfo {author} {\bibfnamefont {T.}~\bibnamefont {Katori}},
  \ and\ \bibinfo {author} {\bibfnamefont {J.}~\bibnamefont {Salvado}},\
  }\href@noop {} {\bibfield  {journal} {\bibinfo  {journal} {Phys. Rev. Lett.}\
  }\textbf {\bibinfo {volume} {115}},\ \bibinfo {pages} {161303} (\bibinfo
  {year} {2015})},\ \Eprint {http://arxiv.org/abs/1506.02043} {arXiv:1506.02043
  [hep-ph]} \BibitemShut {NoStop}%
\bibitem [{\citenamefont {Xing}\ and\ \citenamefont {Zhou}(2008)}]{20}%
  \BibitemOpen
  \bibfield  {author} {\bibinfo {author} {\bibfnamefont {Z.-z.}\ \bibnamefont
  {Xing}}\ and\ \bibinfo {author} {\bibfnamefont {S.}~\bibnamefont {Zhou}},\
  }\href@noop {} {\bibfield  {journal} {\bibinfo  {journal} {Phys. Lett. B}\
  }\textbf {\bibinfo {volume} {666}},\ \bibinfo {pages} {166} (\bibinfo {year}
  {2008})},\ \Eprint {http://arxiv.org/abs/0804.3512} {arXiv:0804.3512
  [hep-ph]} \BibitemShut {NoStop}%
\bibitem [{\citenamefont {Brdar}\ \emph {et~al.}(2017)\citenamefont {Brdar},
  \citenamefont {Kopp},\ and\ \citenamefont {Wang}}]{21}%
  \BibitemOpen
  \bibfield  {author} {\bibinfo {author} {\bibfnamefont {V.}~\bibnamefont
  {Brdar}}, \bibinfo {author} {\bibfnamefont {J.}~\bibnamefont {Kopp}}, \ and\
  \bibinfo {author} {\bibfnamefont {X.-P.}\ \bibnamefont {Wang}},\ }\href@noop
  {} {\bibfield  {journal} {\bibinfo  {journal} {JCAP}\ }\textbf {\bibinfo
  {volume} {01}},\ \bibinfo {pages} {026} (\bibinfo {year} {2017})},\ \Eprint
  {http://arxiv.org/abs/1611.04598} {arXiv:1611.04598 [hep-ph]} \BibitemShut
  {NoStop}%
\bibitem [{\citenamefont {Bustamante}\ \emph {et~al.}(2015)\citenamefont
  {Bustamante}, \citenamefont {Beacom},\ and\ \citenamefont {Winter}}]{22}%
  \BibitemOpen
  \bibfield  {author} {\bibinfo {author} {\bibfnamefont {M.}~\bibnamefont
  {Bustamante}}, \bibinfo {author} {\bibfnamefont {J.~F.}\ \bibnamefont
  {Beacom}}, \ and\ \bibinfo {author} {\bibfnamefont {W.}~\bibnamefont
  {Winter}},\ }\href@noop {} {\bibfield  {journal} {\bibinfo  {journal} {Phys.
  Rev. Lett.}\ }\textbf {\bibinfo {volume} {115}},\ \bibinfo {pages} {161302}
  (\bibinfo {year} {2015})},\ \Eprint {http://arxiv.org/abs/1506.02645}
  {arXiv:1506.02645 [astro-ph.HE]} \BibitemShut {NoStop}%
\bibitem [{\citenamefont {Colladay}\ and\ \citenamefont
  {Kostelecky}(1998)}]{23}%
  \BibitemOpen
  \bibfield  {author} {\bibinfo {author} {\bibfnamefont {D.}~\bibnamefont
  {Colladay}}\ and\ \bibinfo {author} {\bibfnamefont {V.~A.}\ \bibnamefont
  {Kostelecky}},\ }\href@noop {} {\bibfield  {journal} {\bibinfo  {journal}
  {Phys. Rev. D}\ }\textbf {\bibinfo {volume} {58}},\ \bibinfo {pages} {116002}
  (\bibinfo {year} {1998})},\ \Eprint {http://arxiv.org/abs/hep-ph/9809521}
  {arXiv:hep-ph/9809521} \BibitemShut {NoStop}%
\bibitem [{\citenamefont {Coleman}\ and\ \citenamefont {Glashow}(1999)}]{24}%
  \BibitemOpen
  \bibfield  {author} {\bibinfo {author} {\bibfnamefont {S.~R.}\ \bibnamefont
  {Coleman}}\ and\ \bibinfo {author} {\bibfnamefont {S.~L.}\ \bibnamefont
  {Glashow}},\ }\href@noop {} {\bibfield  {journal} {\bibinfo  {journal} {Phys.
  Rev. D}\ }\textbf {\bibinfo {volume} {59}},\ \bibinfo {pages} {116008}
  (\bibinfo {year} {1999})},\ \Eprint {http://arxiv.org/abs/hep-ph/9812418}
  {arXiv:hep-ph/9812418} \BibitemShut {NoStop}%
\bibitem [{\citenamefont {Kostelecky}\ and\ \citenamefont {Mewes}(2004)}]{25}%
  \BibitemOpen
  \bibfield  {author} {\bibinfo {author} {\bibfnamefont {V.~A.}\ \bibnamefont
  {Kostelecky}}\ and\ \bibinfo {author} {\bibfnamefont {M.}~\bibnamefont
  {Mewes}},\ }\href@noop {} {\bibfield  {journal} {\bibinfo  {journal} {Phys.
  Rev. D}\ }\textbf {\bibinfo {volume} {69}},\ \bibinfo {pages} {016005}
  (\bibinfo {year} {2004})},\ \Eprint {http://arxiv.org/abs/hep-ph/0309025}
  {arXiv:hep-ph/0309025} \BibitemShut {NoStop}%
\bibitem [{\citenamefont {Kostelecky}\ and\ \citenamefont
  {Russell}(2011)}]{26}%
  \BibitemOpen
  \bibfield  {author} {\bibinfo {author} {\bibfnamefont {V.~A.}\ \bibnamefont
  {Kostelecky}}\ and\ \bibinfo {author} {\bibfnamefont {N.}~\bibnamefont
  {Russell}},\ }\href@noop {} {\bibfield  {journal} {\bibinfo  {journal} {Rev.
  Mod. Phys.}\ }\textbf {\bibinfo {volume} {83}},\ \bibinfo {pages} {11}
  (\bibinfo {year} {2011})},\ \Eprint {http://arxiv.org/abs/0801.0287}
  {arXiv:0801.0287 [hep-ph]} \BibitemShut {NoStop}%
\bibitem [{\citenamefont {Liberati}(2013)}]{27}%
  \BibitemOpen
  \bibfield  {author} {\bibinfo {author} {\bibfnamefont {S.}~\bibnamefont
  {Liberati}},\ }\href@noop {} {\bibfield  {journal} {\bibinfo  {journal}
  {Class. Quant. Grav.}\ }\textbf {\bibinfo {volume} {30}},\ \bibinfo {pages}
  {133001} (\bibinfo {year} {2013})},\ \Eprint {http://arxiv.org/abs/1304.5795}
  {arXiv:1304.5795 [gr-qc]} \BibitemShut {NoStop}%
\bibitem [{\citenamefont {Boncioli}(2018)}]{28}%
  \BibitemOpen
  \bibfield  {author} {\bibinfo {author} {\bibfnamefont {D.}~\bibnamefont
  {Boncioli}} (\bibinfo {collaboration} {Pierre Auger}),\ }\href@noop {}
  {\bibfield  {journal} {\bibinfo  {journal} {PoS}\ }\textbf {\bibinfo {volume}
  {ICRC2017}},\ \bibinfo {pages} {561} (\bibinfo {year} {2018})}\BibitemShut
  {NoStop}%
\bibitem [{\citenamefont {Lang}\ \emph {et~al.}(2019)\citenamefont {Lang},
  \citenamefont {Mart\'\i{}nez-Huerta},\ and\ \citenamefont {de~Souza}}]{29}%
  \BibitemOpen
  \bibfield  {author} {\bibinfo {author} {\bibfnamefont {R.~G.}\ \bibnamefont
  {Lang}}, \bibinfo {author} {\bibfnamefont {H.}~\bibnamefont
  {Mart\'\i{}nez-Huerta}}, \ and\ \bibinfo {author} {\bibfnamefont
  {V.}~\bibnamefont {de~Souza}},\ }\href@noop {} {\bibfield  {journal}
  {\bibinfo  {journal} {Phys. Rev. D}\ }\textbf {\bibinfo {volume} {99}},\
  \bibinfo {pages} {043015} (\bibinfo {year} {2019})},\ \Eprint
  {http://arxiv.org/abs/1810.13215} {arXiv:1810.13215 [astro-ph.HE]}
  \BibitemShut {NoStop}%
\bibitem [{\citenamefont {Roberts}(2021)}]{30}%
  \BibitemOpen
  \bibfield  {author} {\bibinfo {author} {\bibfnamefont {A.}~\bibnamefont
  {Roberts}},\ }\href@noop {} {\bibfield  {journal} {\bibinfo  {journal}
  {Galaxies}\ }\textbf {\bibinfo {volume} {9}},\ \bibinfo {pages} {47}
  (\bibinfo {year} {2021})}\BibitemShut {NoStop}%
\bibitem [{\citenamefont {Abreu}\ \emph {et~al.}(2022)\citenamefont {Abreu}
  \emph {et~al.}}]{31}%
  \BibitemOpen
  \bibfield  {author} {\bibinfo {author} {\bibfnamefont {P.}~\bibnamefont
  {Abreu}} \emph {et~al.} (\bibinfo {collaboration} {Pierre Auger}),\
  }\href@noop {} {\bibfield  {journal} {\bibinfo  {journal} {JCAP}\ }\textbf
  {\bibinfo {volume} {01}},\ \bibinfo {pages} {023} (\bibinfo {year} {2022})},\
  \Eprint {http://arxiv.org/abs/2112.06773} {arXiv:2112.06773 [astro-ph.HE]}
  \BibitemShut {NoStop}%
\bibitem [{\citenamefont {Zhang}\ and\ \citenamefont {Yang}(2022)}]{32}%
  \BibitemOpen
  \bibfield  {author} {\bibinfo {author} {\bibfnamefont {H.}~\bibnamefont
  {Zhang}}\ and\ \bibinfo {author} {\bibfnamefont {L.}~\bibnamefont {Yang}},\
  }\href@noop {} {\bibfield  {journal} {\bibinfo  {journal} {Universe}\
  }\textbf {\bibinfo {volume} {8}},\ \bibinfo {pages} {260} (\bibinfo {year}
  {2022})}\BibitemShut {NoStop}%
\bibitem [{\citenamefont {Abe}\ \emph {et~al.}(2015)\citenamefont {Abe} \emph
  {et~al.}}]{33}%
  \BibitemOpen
  \bibfield  {author} {\bibinfo {author} {\bibfnamefont {K.}~\bibnamefont
  {Abe}} \emph {et~al.} (\bibinfo {collaboration} {Super-Kamiokande}),\
  }\href@noop {} {\bibfield  {journal} {\bibinfo  {journal} {Phys. Rev. D}\
  }\textbf {\bibinfo {volume} {91}},\ \bibinfo {pages} {052003} (\bibinfo
  {year} {2015})},\ \Eprint {http://arxiv.org/abs/1410.4267} {arXiv:1410.4267
  [hep-ex]} \BibitemShut {NoStop}%
\bibitem [{\citenamefont {Beringer}\ \emph {et~al.}(2012)\citenamefont
  {Beringer} \emph {et~al.}}]{34}%
  \BibitemOpen
  \bibfield  {author} {\bibinfo {author} {\bibfnamefont {J.}~\bibnamefont
  {Beringer}} \emph {et~al.} (\bibinfo {collaboration} {Particle Data Group}),\
  }\href@noop {} {\bibfield  {journal} {\bibinfo  {journal} {Phys. Rev. D}\
  }\textbf {\bibinfo {volume} {86}},\ \bibinfo {pages} {010001} (\bibinfo
  {year} {2012})}\BibitemShut {NoStop}%
\bibitem [{\citenamefont {Aartsen}\ \emph
  {et~al.}(2018{\natexlab{a}})\citenamefont {Aartsen} \emph {et~al.}}]{35}%
  \BibitemOpen
  \bibfield  {author} {\bibinfo {author} {\bibfnamefont {M.~G.}\ \bibnamefont
  {Aartsen}} \emph {et~al.} (\bibinfo {collaboration} {IceCube}),\ }\href@noop
  {} {\bibfield  {journal} {\bibinfo  {journal} {Nature Phys.}\ }\textbf
  {\bibinfo {volume} {14}},\ \bibinfo {pages} {961} (\bibinfo {year}
  {2018}{\natexlab{a}})},\ \Eprint {http://arxiv.org/abs/1709.03434}
  {arXiv:1709.03434 [hep-ex]} \BibitemShut {NoStop}%
\bibitem [{\citenamefont {Esteban}\ \emph {et~al.}(2020)\citenamefont
  {Esteban}, \citenamefont {Gonzalez-Garcia}, \citenamefont {Maltoni},
  \citenamefont {Schwetz},\ and\ \citenamefont {Zhou}}]{36}%
  \BibitemOpen
  \bibfield  {author} {\bibinfo {author} {\bibfnamefont {I.}~\bibnamefont
  {Esteban}}, \bibinfo {author} {\bibfnamefont {M.~C.}\ \bibnamefont
  {Gonzalez-Garcia}}, \bibinfo {author} {\bibfnamefont {M.}~\bibnamefont
  {Maltoni}}, \bibinfo {author} {\bibfnamefont {T.}~\bibnamefont {Schwetz}}, \
  and\ \bibinfo {author} {\bibfnamefont {A.}~\bibnamefont {Zhou}},\ }\href@noop
  {} {\bibfield  {journal} {\bibinfo  {journal} {JHEP}\ }\textbf {\bibinfo
  {volume} {09}},\ \bibinfo {pages} {178} (\bibinfo {year} {2020})},\ \Eprint
  {http://arxiv.org/abs/2007.14792} {arXiv:2007.14792 [hep-ph]} \BibitemShut
  {NoStop}%
\bibitem [{\citenamefont {Aartsen}\ \emph
  {et~al.}(2018{\natexlab{b}})\citenamefont {Aartsen} \emph {et~al.}}]{37}%
  \BibitemOpen
  \bibfield  {author} {\bibinfo {author} {\bibfnamefont {M.~G.}\ \bibnamefont
  {Aartsen}} \emph {et~al.} (\bibinfo {collaboration} {IceCube, Fermi-LAT,
  MAGIC, AGILE, ASAS-SN, HAWC, H.E.S.S., INTEGRAL, Kanata, Kiso, Kapteyn,
  Liverpool Telescope, Subaru, Swift NuSTAR, VERITAS, VLA/17B-403}),\
  }\href@noop {} {\bibfield  {journal} {\bibinfo  {journal} {Science}\ }\textbf
  {\bibinfo {volume} {361}},\ \bibinfo {pages} {eaat1378} (\bibinfo {year}
  {2018}{\natexlab{b}})},\ \Eprint {http://arxiv.org/abs/1807.08816}
  {arXiv:1807.08816 [astro-ph.HE]} \BibitemShut {NoStop}%
\bibitem [{\citenamefont {Abbasi}\ \emph {et~al.}(2021)\citenamefont {Abbasi}
  \emph {et~al.}}]{38}%
  \BibitemOpen
  \bibfield  {author} {\bibinfo {author} {\bibfnamefont {R.}~\bibnamefont
  {Abbasi}} \emph {et~al.} (\bibinfo {collaboration} {IceCube}),\ }\href@noop
  {} {\bibfield  {journal} {\bibinfo  {journal} {Phys. Rev. D}\ }\textbf
  {\bibinfo {volume} {104}},\ \bibinfo {pages} {022002} (\bibinfo {year}
  {2021})},\ \Eprint {http://arxiv.org/abs/2011.03545} {arXiv:2011.03545
  [astro-ph.HE]} \BibitemShut {NoStop}%
\bibitem [{\citenamefont {Anchordoqui}\ \emph
  {et~al.}(2014{\natexlab{a}})\citenamefont {Anchordoqui}, \citenamefont
  {Goldberg}, \citenamefont {Lynch}, \citenamefont {Olinto}, \citenamefont
  {Paul},\ and\ \citenamefont {Weiler}}]{39}%
  \BibitemOpen
  \bibfield  {author} {\bibinfo {author} {\bibfnamefont {L.~A.}\ \bibnamefont
  {Anchordoqui}}, \bibinfo {author} {\bibfnamefont {H.}~\bibnamefont
  {Goldberg}}, \bibinfo {author} {\bibfnamefont {M.~H.}\ \bibnamefont {Lynch}},
  \bibinfo {author} {\bibfnamefont {A.~V.}\ \bibnamefont {Olinto}}, \bibinfo
  {author} {\bibfnamefont {T.~C.}\ \bibnamefont {Paul}}, \ and\ \bibinfo
  {author} {\bibfnamefont {T.~J.}\ \bibnamefont {Weiler}},\ }\href@noop {}
  {\bibfield  {journal} {\bibinfo  {journal} {Phys. Rev. D}\ }\textbf {\bibinfo
  {volume} {89}},\ \bibinfo {pages} {083003} (\bibinfo {year}
  {2014}{\natexlab{a}})},\ \Eprint {http://arxiv.org/abs/1306.5021}
  {arXiv:1306.5021 [astro-ph.HE]} \BibitemShut {NoStop}%
\bibitem [{\citenamefont {Anchordoqui}\ \emph
  {et~al.}(2014{\natexlab{b}})\citenamefont {Anchordoqui} \emph {et~al.}}]{40}%
  \BibitemOpen
  \bibfield  {author} {\bibinfo {author} {\bibfnamefont {L.~A.}\ \bibnamefont
  {Anchordoqui}} \emph {et~al.},\ }\href@noop {} {\bibfield  {journal}
  {\bibinfo  {journal} {JHEAp}\ }\textbf {\bibinfo {volume} {1-2}},\ \bibinfo
  {pages} {1} (\bibinfo {year} {2014}{\natexlab{b}})},\ \Eprint
  {http://arxiv.org/abs/1312.6587} {arXiv:1312.6587 [astro-ph.HE]} \BibitemShut
  {NoStop}%
\bibitem [{\citenamefont {Murase}\ and\ \citenamefont {Stecker}(2022)}]{41}%
  \BibitemOpen
  \bibfield  {author} {\bibinfo {author} {\bibfnamefont {K.}~\bibnamefont
  {Murase}}\ and\ \bibinfo {author} {\bibfnamefont {F.~W.}\ \bibnamefont
  {Stecker}},\ }\href@noop {} {\  (\bibinfo {year} {2022})},\ \Eprint
  {http://arxiv.org/abs/2202.03381} {arXiv:2202.03381 [astro-ph.HE]}
  \BibitemShut {NoStop}%
\bibitem [{\citenamefont {Sironi}\ \emph {et~al.}(2015)\citenamefont {Sironi},
  \citenamefont {Petropoulou},\ and\ \citenamefont {Giannios}}]{42}%
  \BibitemOpen
  \bibfield  {author} {\bibinfo {author} {\bibfnamefont {L.}~\bibnamefont
  {Sironi}}, \bibinfo {author} {\bibfnamefont {M.}~\bibnamefont {Petropoulou}},
  \ and\ \bibinfo {author} {\bibfnamefont {D.}~\bibnamefont {Giannios}},\
  }\href@noop {} {\bibfield  {journal} {\bibinfo  {journal} {Mon. Not. Roy.
  Astron. Soc.}\ }\textbf {\bibinfo {volume} {450}},\ \bibinfo {pages} {183}
  (\bibinfo {year} {2015})},\ \Eprint {http://arxiv.org/abs/1502.01021}
  {arXiv:1502.01021 [astro-ph.HE]} \BibitemShut {NoStop}%
\bibitem [{\citenamefont {Petropoulou}\ \emph {et~al.}(2016)\citenamefont
  {Petropoulou}, \citenamefont {Giannios},\ and\ \citenamefont {Sironi}}]{43}%
  \BibitemOpen
  \bibfield  {author} {\bibinfo {author} {\bibfnamefont {M.}~\bibnamefont
  {Petropoulou}}, \bibinfo {author} {\bibfnamefont {D.}~\bibnamefont
  {Giannios}}, \ and\ \bibinfo {author} {\bibfnamefont {L.}~\bibnamefont
  {Sironi}},\ }\href@noop {} {\bibfield  {journal} {\bibinfo  {journal} {Mon.
  Not. Roy. Astron. Soc.}\ }\textbf {\bibinfo {volume} {462}},\ \bibinfo
  {pages} {3325} (\bibinfo {year} {2016})},\ \Eprint
  {http://arxiv.org/abs/1606.07447} {arXiv:1606.07447 [astro-ph.HE]}
  \BibitemShut {NoStop}%
\bibitem [{\citenamefont {French}\ \emph {et~al.}(2022)\citenamefont {French},
  \citenamefont {Guo}, \citenamefont {Zhang},\ and\ \citenamefont
  {Uzdensky}}]{44}%
  \BibitemOpen
  \bibfield  {author} {\bibinfo {author} {\bibfnamefont {O.}~\bibnamefont
  {French}}, \bibinfo {author} {\bibfnamefont {F.}~\bibnamefont {Guo}},
  \bibinfo {author} {\bibfnamefont {Q.}~\bibnamefont {Zhang}}, \ and\ \bibinfo
  {author} {\bibfnamefont {D.}~\bibnamefont {Uzdensky}},\ }\href@noop {} {\
  (\bibinfo {year} {2022})},\ \Eprint {http://arxiv.org/abs/2210.08358}
  {arXiv:2210.08358 [astro-ph.HE]} \BibitemShut {NoStop}%
\end{thebibliography}%

\end{document}